\documentclass[preprint,floats,floatfix,aps,epsfig,nofootinbib,amssymb]{revtex4}
\usepackage{subfigure}
\usepackage{graphics,psfrag}
\usepackage{graphicx,psfrag}
\usepackage{amsmath,amssymb}
\usepackage[sort&compress]{natbib}
\usepackage{cancel}
\usepackage{verbatim}
\usepackage{multirow}
\usepackage{bm}
\def\invfb{{\rm fb}^{-1}}
\def\afbt{A_{FB}^{\ttb}}
\def\gev{\rm GeV}
\def\tev{\rm TeV}

\def\ttb{t\bar t}
\def\etmiss{{E\!\!\!\!\slash_{T}}}

\def\lsim{\mathrel{\raise.3ex\hbox{$<$\kern-.75em\lower1ex\hbox{$\sim$}}}}
\def\gsim{\mathrel{\raise.3ex\hbox{$>$\kern-.75em\lower1ex\hbox{$\sim$}}}}
\def\ifmath#1{\relax\ifmmode #1\else $#1$\fi}
\def\beq{\begin{equation}}
\def\eeq{\end{equation}}
\def\bea{\begin{eqnarray}}
\def\eea{\end{eqnarray}}
\def\nn{\nonumber}
\def\bc{}

\begin{document}
\draft

\preprint{
   {\vbox {
      \hbox{\bf MSUHEP-120607}
      }}}
\vspace*{2cm}

\title{Probing Color Octet Couplings at the Large Hadron Collider}
\vspace*{0.25in}   
\author{{Anupama Atre, R. Sekhar Chivukula, Pawin Ittisamai, \\ 
Elizabeth H. Simmons and Jiang-Hao Yu}
\footnote{avatre@pa.msu.edu\\ sekhar@pa.msu.edu\\ ittisama@msu.edu\\ esimmons@pa.msu.edu\\ yujiangh@msu.edu.} }  
\affiliation{\vspace*{0.1in}
Department of Physics and Astronomy\\
Michigan State University, East Lansing U.S.A.\\}
\vspace*{0.25 in} 

\begin{abstract}
\vspace{0.5cm}
Color-octet resonances arise in many well motivated theories beyond the standard model. As colored objects they are produced copiously at the LHC and can be discovered in early searches for new physics in dijet final states. Once they are discovered it will be important to measure the couplings of the new resonances to determine the underlying theoretical structure. We propose a new channel, associated production of $W,Z$ gauge bosons and color-octet resonances, to help determine the chiral structure of the couplings. We present our analysis for a range of color-octet masses (2.5 to 4.5 TeV), couplings and decay widths for the LHC with center of mass energy of 14 TeV and 10 $\invfb$ or 100 $\invfb$ of integrated luminosity. We find that the LHC can probe a large region of the parameter space up to very small couplings. \\

\end{abstract} 

\maketitle

\section{Introduction}
\label{sec:introd}

The CERN Large Hadron Collider (LHC) has been successfully accumulating data at the highest recorded collision energies, opening up the teraelectron-volt (TeV) energy region to direct experimental exploration. At a hadron collider as the initial states are composed of colored particles, the LHC will produce new colored resonances at a large rate. This makes the LHC an ideal machine for exploring new colored particles associated with new strong dynamics.

There are many theories of physics beyond the Standard Model (BSM) that give rise to new colored particles. One such possibility is a massive vector boson in the color-octet representation.  Such a particle has been proposed in various scenarios of which we list a sample here. Colorons are predicted in the flavor-universal coloron model, in which the QCD gauge group is extended to $SU(3)_1 \times SU(3)_2$ with all the quarks assigned to the triplet representation of the strong $SU(3)_2$ group~\cite{Chivukula:1996yr, Simmons:1996fz}. Topgluons represent a coloron variant in which the quarks of the third generation are assigned to a representation of one $SU(3)$ group and the light quarks are assigned to the other; they arise in topcolor models~\cite{Hill:1991at, Hill:1994hp}. Axigluons appear in the chiral color model, in which the extension of QCD gauge structure is through new strong chiral gauge symmetry $SU(3)_L \times SU(3)_R$ and can feature several kinds of assignment of quark charges to the gauge groups~\cite{Frampton:1987dn, Frampton:1987ut, Bagger:1987fz, Chivukula:2010fk, Frampton:2009rk}. Other examples include Kaluza-Klein (KK) gluons which are the excited gluons in extra-dimensional models \cite{Dicus:2000hm}, technirhos which are composite colored vector mesons found in Technicolor~\cite{Farhi:1980xs,Hill:2002ap, Lane:2002sm}, models that include colored technifermions and low-scale string resonances ~\cite{Antoniadis:1990ew}.

The high production rate of a colored resonance (due to strong coupling) and the simple topology of the final state (decay into two jets) makes the search for di-jet resonances one of the early signatures that are studied at hadron colliders, e.g. the CERN $\rm{S\bar{p}pS}$~\cite{Arnison:1986vk, Alitti:1993pn}, Tevatron~\cite{Abe:1989gz, Abe:1995jz, Abe:1997hm, Abazov:2003tj, Aaltonen:2008dn} and LHC~\cite{Aad:2010ae, Khachatryan:2010jd, Chatrchyan:2011ns, Aad:2011fq, atlas:2012-038}. Once a new colored resonance is discovered, measuring its properties will be the next important task. The di-jet invariant mass $m_{jj}$ and the angular distributions of energetic jets relative to the beam axis are sensitive observables to determine the properties, such as mass and spin of the resonance. Although one can constrain the coupling strength of the colored resonance to the Standard Model (SM) quarks using the total cross section, this is not sufficient to determine the chiral structure of the couplings. 

Some additional information about the chiral couplings can potentially be gleaned from the measurement of the forward-backward asymmetry of the top-antitop pair ($\afbt$). Such a measurement has been made at the Tevatron \cite{Aaltonen:2011kc, Abazov:2011rq}. Motivated by these results, various models with specific chiral structures of the color-octet particle that give rise to a large asymmetry have been proposed ~\cite{Chivukula:2010fk}\cite{Frampton:2009rk}. Unfortunately due to the proton-proton initial states at the LHC, the forward backward asymmetry of the top-antitop pair is diluted  at that machine compared to the large asymmetry at the Tevatron \cite{Chatrchyan:2011hk,ATLAS:2012an}. 

\begin{figure}[tb]
{\includegraphics[width=0.50\textwidth,clip=true]{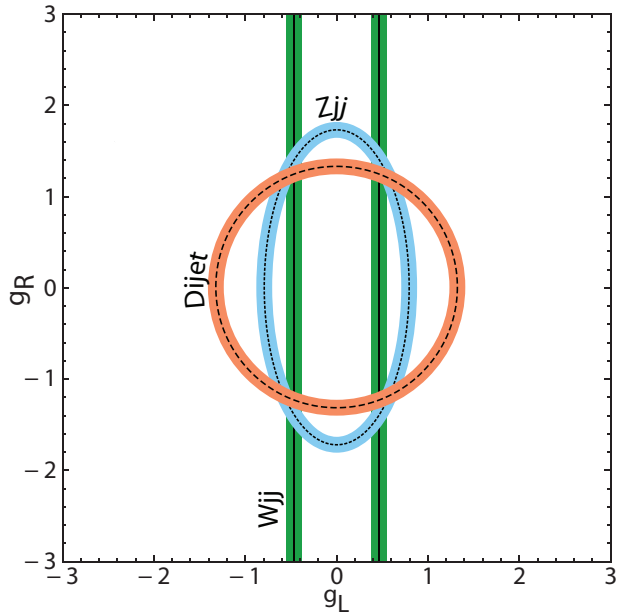}}
\caption{ 
A cartoon illustration of the form of prospective constraints on chiral couplings from the di-jet channel (dashed black circle with red band), the channel with associated production of a $W$ boson (solid black parallel lines with green band) and the channel with associated production of a $Z$ boson (dotted black ellipse with blue band). Combining the constraints from different channels will narrow the range of allowed couplings.}
\label{fig:cartooncouplings}
\end{figure}

We propose a new channel for studying coloron couplings: the associated production of a $W$ or $Z$ gauge boson with the color-octet at the LHC. The chiral couplings of the weak gauge bosons to the fermions in the associated production channel provides additional information about the chiral structure of the new strong dynamics. Combining the associated production channel with the di-jet channel makes it possible to extract the chiral couplings of the colored resonance because the cross-sections of each channel have a different dependence on the coloron's couplings to fermions. The functional form of the dependance of these measurements on the chiral couplings in the di-jet channel is $ g^2_L + g^2_R$; in the Wjj channel it is $g^2_L$ and in the Zjj channel it is $ a g^2_L + b g^2_R$. A cartoon illustration of these three measurements along with the di-jet measurement is shown in Fig.~\ref{fig:cartooncouplings}. Notice that while combining the different channels will narrow the allowed range of couplings, there remains an  ambiguity in extracting the sign of the couplings. This method of using the associated production of a weak gauge boson to illuminate the properties of a new resonance was studied earlier in the context of the measurement of $Z^\prime$ couplings \cite{Cvetic:1992qv,delAguila:1993ym}. 

In this article we study the sensitivity of the LHC with center of mass (c.m.) energy of 14 TeV to probe the chiral structure of the couplings for colored resonances with 10 $\invfb$ and 100 $\invfb$ integrated luminosity by the method proposed above. We study colored resonances with masses in the range 2.5 TeV to 4.5 TeV and various couplings and widths;  this mass range runs from the lightest colorons still allowed by LHC dijet searches to the heaviest colorons to which the LHC is likely to be sensitive. The rest of the paper is organized as follows. In Sec.\ref{sec:genpar} we present a simple parameterization for the colored resonances and our notation. In Sec.\ref{sec:collph} we discuss the signal and associated backgrounds, the Monte Carlo simulation details in Sec.\ref{sec:collph-mcsim} and the channels with charged and neutral gauge bosons in Sec.~\ref{sec:collph-cw} and Sec.\ref{sec:collph-cz} respectively. We present a discussion of our results in Sec.\ref{sec:sensitivity} and conclusions in Sec.\ref{sec:concl}.

\section{General Parameterization}
\label{sec:genpar}

The color-octet resonance of interest to our study may be motivated in many BSM scenarios as noted in the introduction. Hence we explore a phenomenological model of color-octet resonances independent of the underlying theory. The interaction of the color-octet resonance $C^\mu$ with the SM quarks $q_i$ has the form
\beq
\mathcal{L} = i g_s\bar{q}_iC^\mu\gamma_\mu\left(g_V^i + g_A^i\gamma_5\right) q_i = i g_s\bar{q}_iC^\mu\gamma_\mu\left( g_L^i P_L + g_R^i P_R \right) q_i, 
\label{eq:axicoupl}
\eeq
where $C_\mu = C_\mu^a t^a$ with $t^a$ an $ \text{SU}(3) $ generator, $g_V^i$ and $g_A^i$ (or $g_L^i$ and $g_R^i$) denote vector and axial (or left and right) coupling strengths relative to the QCD coupling $g_s$, the projection operators have the form $P_{L,R} = (1 \mp \gamma_5)/2$ and the quark flavors run over $ i=u,c,d,s,b,t.$ For simplicity, we will denote the color-octet resonance by $C$ and its chiral couplings to light quarks by $g_{L,R}^q$ and to the third generation by $g_{L,R}^t$. 

The various couplings $g_{L,R}^{q,t}$ can all be independent in the most general non-universal case and can all be the same in simple flavor universal scenarios. A study of the former case would be desirable but it is computationally cumbersome and the latter case, while simple, does not include some interesting scenarios. Furthermore the authors of Ref.~\cite{Haisch:2011up} found that the flavor universal case is excluded at 99.5\% by a global fit to top pair data for color-octet masses less than 3 TeV. An interesting (yet manageable and general) example is that of a color-octet resonance that can enhance top-pair forward-backward asymmetry in models allowing non-universal couplings when \cite{Ferrario:2009bz, Ferrario:2008wm, Rodrigo:2010gm}
\beq
g_A^q  g_A^t < 0.
\label{eq:gaqgat}
\eeq
Motivated by the $\afbt$ measurement at the Tevatron we choose the following non-universal couplings as in Ref.~\cite{Ferrario:2009bz} as an example. %
\beq
g_V^t = g_V^q	\qquad\mathrm{and}\qquad g_A^t = -g_A^q,
\label{eq:gaqgatrel}
\eeq
or in terms of $g_L-g_R$, 
\beq
g_L^t = g_R^q, \qquad\text{and}\qquad g_R^t = g_L^q.
\label{eq:glqgrtrel}
\eeq
This example is sufficient to demonstrate the utility of the method of measuring couplings described in this article. This choice of couplings also allows us to easily compare the reach of our study with that of Ref.~\cite{Ferrario:2009bz} which is consistent with the $\afbt$ measurement. A more general study involving fully independent couplings is beyond the scope of this study and will be addressed in a future publication. 

The color-octet resonance with the interactions as in Eq.~(\ref{eq:axicoupl}) decays primarily to two jets or a top pair and its decay width is given by
\bea
\nn
\Gamma_C &=& \alpha_s \frac{m_C}{12} \Big[ 4 \left(g_L^{q\ 2} + g_R^{q\ 2}\right) + \left( g_L^{t\ 2} + g_R^{t\ 2} \right) \\
&+& \left[ (g_L^{t\ 2} + g_R^{t\ 2}) (1 -  \mu_t)  + 6g_L^{t}g_R^{t} \mu_t \right] \sqrt{1 - 4 \mu_t} \Big],   
\label{eq:axiwid}
\eea
where the terms in the first line come from decays to light quarks and to the bottom quark, while the terms in the second line come from decays to top quarks. Decays to top quarks are modified by the kinematic factors involving $\mu_{t} = m_{top}^2/m_C^2$ with $m_{top}$ and $m_C$ the top quark and color-octet mass respectively. Strictly speaking, the bottom quark's contribution to the width is modified by factors involving $\mu_b$ but we ignore these factors since $m_b^2 << m_C^2$. For an octet that is heavy compared to the top quark and whose couplings are of the form given in Eq.~(\ref{eq:glqgrtrel}) the expression for the decay width simplifies to
\beq
\Gamma_C = \frac{\alpha_s}{2} m_C \left( g_L^2 + g_R^2\right) = \alpha_s m_C \left( g_V^2 + g_A^2\right) .
\label{eq:axiwidsimple}
\eeq
In this simplified version, the branching fraction for the color-octet resonance to decay to any single quark flavor obeys the simple relation
\beq
BR( C \rightarrow q_i \bar{q}_i ) = 1/6, \hspace{1cm} {\mbox {\text where}\  i=u,d,c,s,t,b}. 
\label{eq:axibr}
\eeq
%

\section{Collider Phenomenology}
\label{sec:collph}

In this section we study the collider phenomenology of color-octet states 
produced in association with a $W$ or $Z$ gauge boson and discuss the signal and associated backgrounds. We present the the Monte Carlo simulation details in Sec.\ref{sec:collph-mcsim}, and in Sec.\ref{sec:collph-cw}  and Sec.\ref{sec:collph-cz} we study the modes of associated production with a $W$ and a $Z$ boson, respectively. 
 
\begin{figure}[tb]
{\includegraphics[width=0.3\textwidth]{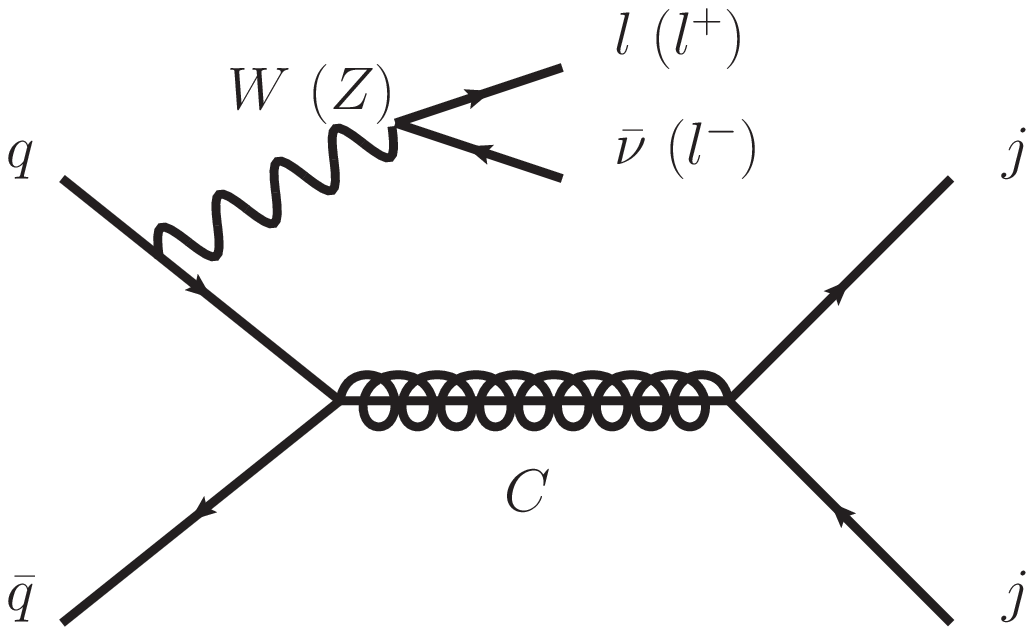} \hspace{1cm}
\includegraphics[width=0.35\textwidth]{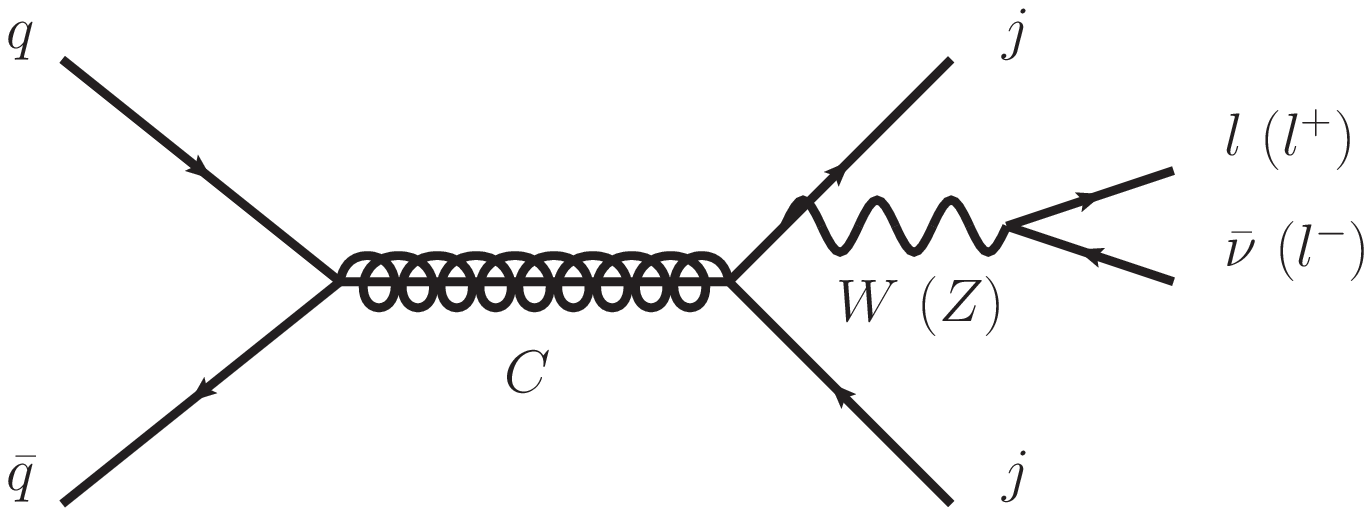} \hspace{1cm}
\includegraphics[width=0.3\textwidth,clip=true]{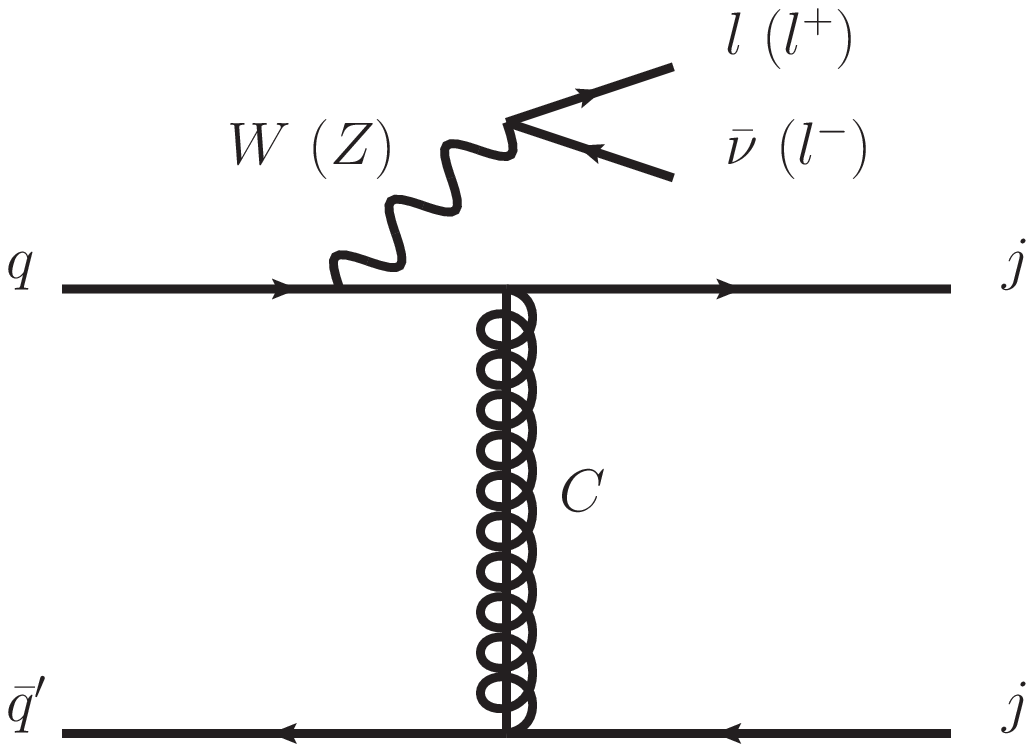}\hspace{1cm}
\includegraphics[width=0.3\textwidth,clip=true]{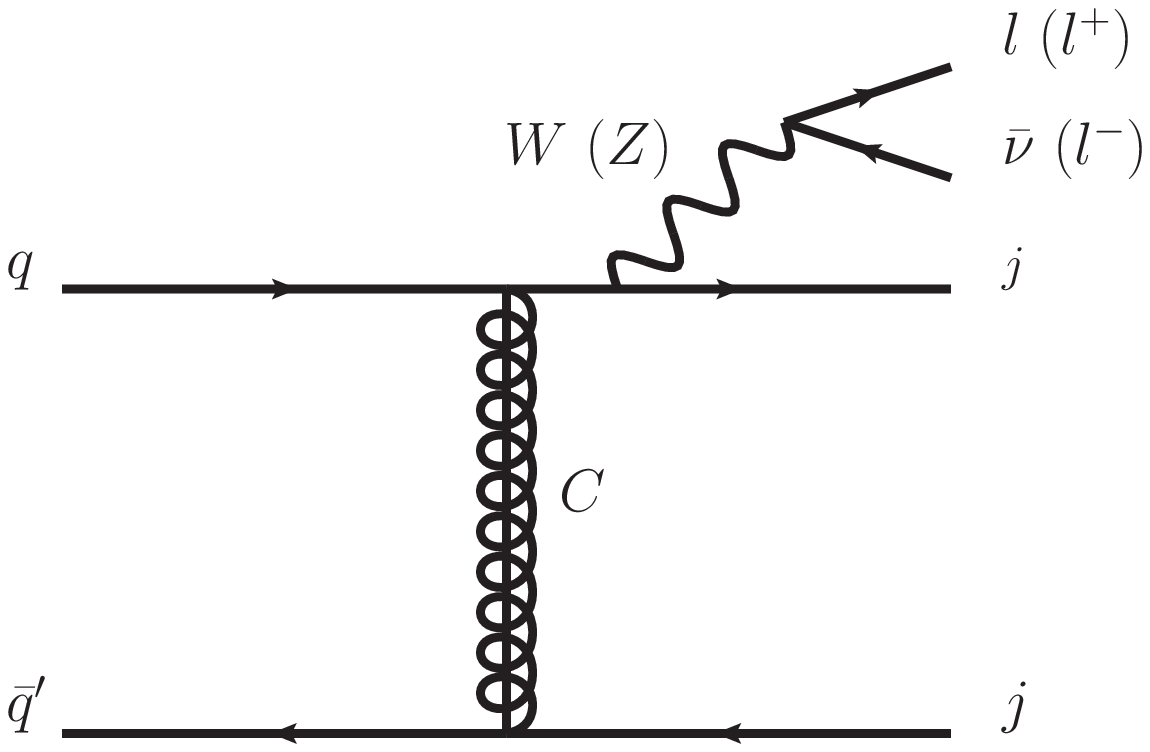}}
\caption{ Representative Feynman diagrams for associated production of a $W,Z$ gauge boson with a color-octet resonance, C. Both $s$ and $t$ channel diagrams along with initial or final state radiation of the associated gauge boson are shown. We assume that the weak gauge boson decays leptonically. }
\label{fig:sigdiag}
\end{figure}

The color-octet states ($C$) are produced and decay to two jets via the process 
\beq
p p \stackrel {C} {\longrightarrow} j\ j. 
\label{eq:prodndijet}
\eeq
They can also be produced in association with a weak gauge boson via the processes	
\bea
\label{eq:cw}
p p  &\stackrel {C} {\longrightarrow}& j\ j\ W^\pm ,  \\
\label{eq:cz}
p p  &\stackrel {C} {\longrightarrow}& j\ j\ Z, 
\eea
where $j=u,d,s,c,b$. We will refer to the processes in Eq.~(\ref{eq:cw}) and Eq.~(\ref{eq:cz}) as the $CW$ and $CZ$ channels, respectively. Representative diagrams of interest for the associated production modes, which include $s$ and $t$ channel diagrams with the emission of the gauge bosons in either the initial or final state, are shown in Fig.~\ref{fig:sigdiag}. The final state channels that we study are
\beq
pp\  \to\  \ell^\pm \etmiss\ 2j,\quad \ell^+\ell^-\ 2j,
\label{eq:sig}
\eeq
coming from $W^\pm(\to \ell^\pm \nu)$ or $Z (\rightarrow \ell^+ \ell^-)$, respectively and $\ell = e, \mu$. Although the inclusion of the $\tau$ lepton in the final state could increase signal statistics, for simplicity we ignore this experimentally more challenging channel. 

The relevant backgrounds to the signal processes in Eq.~(\ref{eq:sig}) are
\begin{itemize}
\item $W+$ jets, $Z+$ jets with $W, Z$ leptonic decays;
\item top pair production with fully leptonic, semi-leptonic and hadronic decays  (where some final state particles may be missed or mis-identified);
\item single top production leading to a $W^\pm b\ q$ final state;
\item $W^+W^-, W^\pm Z$ and $ZZ$ with all possible decay combinations leading to the final state in Eq.~(\ref{eq:sig});
\end{itemize}

Next, we present some details about the Monte Carlo simulation.    
 
\subsection{Monte Carlo Simulation}
\label{sec:collph-mcsim}

We have performed a detailed simulation of both the signal and all the
relevant backgrounds using Madgraph/Madevent \cite{Alwall:2007st} for
event generation at the partonic level, PYTHIA \cite{Sjostrand:2006za} for parton showering with initial and final state radiation as well as hadronization and PGS~\cite{pgs4} for detector simulation. All the detector simulation parameters were set to default values that correspond to the LHC Detector in Madgraph/Madevent.  The CTEQ6L1 Parton Distribution Functions (PDFs)~\cite{Pumplin:2002vw} were used for both signal and background samples. For the signal processes we set the factorization and renormalization scales to be  $\mu_F=\mu_R=m_C$. For the background processes the renormalization and factorization scales are set to $\mu_F = \mu_R = Q$ where $Q^2 = \sum (m_i^2 + p_T^{j_i 2})$. 

We simulate signal samples with color-octet masses, $m_C = 2.5$ to $4.5\  (4.0)\  \tev$ for the  $CW\ (CZ)$ channel. For each mass, we do an exhaustive sampling of couplings $g^q_{L,R}, g^t_{L,R}$, where the couplings $g^t_{L,R}$ follow the relation in Eq.~(\ref{eq:glqgrtrel}). This exhaustive sampling of couplings gives us color-octets with varying widths, $\Gamma_C/m_C \sim 0.025\ -\ 0.50$. While we use the entire sample in estimating the reach of the signal, we will only present details about the samples with $\Gamma_C/m_C = 0.05, 0.10, 0.20, 0.30$ for simplicity and clarity of presentation. We show the parton-level cross sections for a few different color-octet masses and widths at the LHC with $\sqrt{s} = 14\ \tev$ in Table~\ref{tab:xsecs} for illustration.  Also shown, for comparison, are the sizes of the standard model backgrounds at parton level.

\begin{table}[!h]
\begin{tabular}{| c| c| c |c|c | c|c|c | c| c| c|}
\cline{1-8}
\cline{10-11}
\multirow{2}{*}{$m_C$ (TeV)} & \multirow{2}{*}{$\Gamma/m_C$} & \multicolumn{3}{c|}{Signal - $CW$} & \multicolumn{3}{c|}{Signal - $CZ$} & \hspace{1.0cm} &\multirow{2}{*}{Background} & \multirow{2}{*}{$\sigma$ (fb)}	\\
\cline{3-8}
 & & $\ \ g_L^q\ \ $ & $\ \ g_R^q\ \ $ & $\sigma$ (fb) & $\ \ g_L^q\ \ $ & $\ \ g_R^q\ \ $ & $\sigma$ (fb)	& &  & \\
\cline{1-8}
\cline{10-11}
2.5 & 0.05 & -0.42 & 0.82 & 8.4 & -0.82 & -0.42 & 3.5 & &$(W \to \ell \nu)$ + 2 jets &    9500          \\
3.0 & 0.10 & 0.59 & 1.2 & 7.5	 & -1.09 & -0.71 & 3.0 & &$t \bar t$ semi leptonic   &  4200      \\
3.5 &	 0.20 & 0.71 & 1.7 & 6.4	 & -1.49 & 1.08 & 3.5 & &$(Z \to \ell \ell)$ + 2 jets &     1000         \\
4.0 & 0.30 & 1.2 & 1.8 & 8.2 & -1.92 & -1.17 & 3.8  & &single top ($t \to \ell^\pm \nu b$)     & 160  \\
4.5 & 0.30 & -1.2 & 1.9 & 7.8 & - & - & -  & &Total & 15000\\
\cline{1-8}
\cline{10-11}
\end{tabular}
\caption{Representative cross sections at parton level in fb for the $CW$ and $CZ$ signal modes and backgrounds at the LHC with $\sqrt{s} = 14$ TeV. The signal is shown for $m_C = 2.5$ to $4.5\  \tev$ and for a few sample coupling values that correspond to $\Gamma_C/m_C = 0.05,\  0.10,\  0.20$ and $0.30$. For the backgrounds a requirement that $p_T^{j_1}> 250$\,GeV and  $p_T^{j_2}> 200$\,GeV was applied in all cases, except for $t\bar t$ where  $p_T^{j_1}> 150$\,GeV and  $p_T^{j_2}> 100$\,GeV was applied. }
\label{tab:xsecs}
\end{table}

The backgrounds for the process of interest are listed in Sec.\ref{sec:collph}. We generate all the backgrounds with the requirement that $p_T^{j_1}> 250$\,GeV and  $p_T^{j_2}> 200$\,GeV (at the parton level) except top-pair production where we used $p_T^{j_1}> 150$\,GeV and  $p_T^{j_2}> 100$\,GeV. We verified that requiring large cuts on the $p_T$ of the leading jets at the parton level does not distort the distributions in the region of interest to us (higher $p_T$ regions) or affect the background efficiencies. 

For the $CW$ channel the leading background is $W+$ jets with leptonic decays of the $W$. There is also a sizable contribution to the background from top pair production with semi-leptonic decays. Other decay modes of top pair such as the fully leptonic mode where one lepton is lost and the hadronic mode where there is a fake lepton turn out to be negligible after the acceptance and optimized cuts are applied. The single top background as well as those from $Z+$ jets where one lepton is lost turn out to be relevant at the sub-leading level. The background from all the diboson channels turns out to be insignificant after the acceptance and optimized cuts are applied. For the $CZ$ channel the only relevant background is $Z+$ jets with leptonic decays of $Z$. The other backgrounds such as top pair with fully leptonic decays and diboson channels turn out to be insignificant after the acceptance and optimized cuts  are included. While we analyzed all the backgrounds, we only list the cross sections for the  leading and sub-leading backgrounds in Table~\ref{tab:xsecs} for illustration. 

We summarize below the minimum cuts  imposed (after detector simulation) in reconstructing the physics objects  for our analysis. 
\bea
\nn
&& p_T^j > 40\ {\gev}, \quad |\eta_j | < 2.5, \quad \Delta R(jj) > 0.4, \quad \#\  {\mbox {\text jets}} \ge 2, \\
\nn
&& p_T^\ell > 25\ {\gev}, \quad |\eta_\ell| < 2.5,\quad \Delta R(j\ell) > 0.4, \quad \Delta R(\ell\ell) > 0.2,\\
\nn
&&  \etmiss > 25\ {\gev}, \quad \# \ {\mbox {\text isolated\  leptons}} = 1 (2) \ {\mbox {\text for\ }} CW (CZ)\  {\mbox {\text channel,}}\\
&& m_Z - 20\  {\gev} < m_{\ell\ell} < m_Z + 20\  {\gev}.
\label{eq:basiccuts}
\eea
Note that the $\etmiss$ cut is only for the $CW$ channel and the $\Delta R(\ell\ell)$ and $m_{\ell\ell}$ cuts are only for the CZ channel. All other cuts are applied for both channels. 
 
\subsection{Charged Current Channel with $W$-Decays to Leptons }
\label{sec:collph-cw}

In this section we study the case where the color-octet is produced in association with a $W$ boson and investigate the prospects of this channel at the LHC with $\sqrt{s} = 14\  \tev$. The color-octet decays to two jets and the $W$  boson decays leptonically $W\to \ell \nu\ (\ell=e,\mu)$ giving us a final state $jj\ell\etmiss$. We probe color-octet masses from 2.5 to 4.5 TeV for a wide range of couplings $g^q_{L,R}, g^t_{L,R}$ and decay widths ($\Gamma_C$) and describe the analysis in detail below. 

\begin{figure}[tb]
{\includegraphics[width=0.495\textwidth,clip=true]{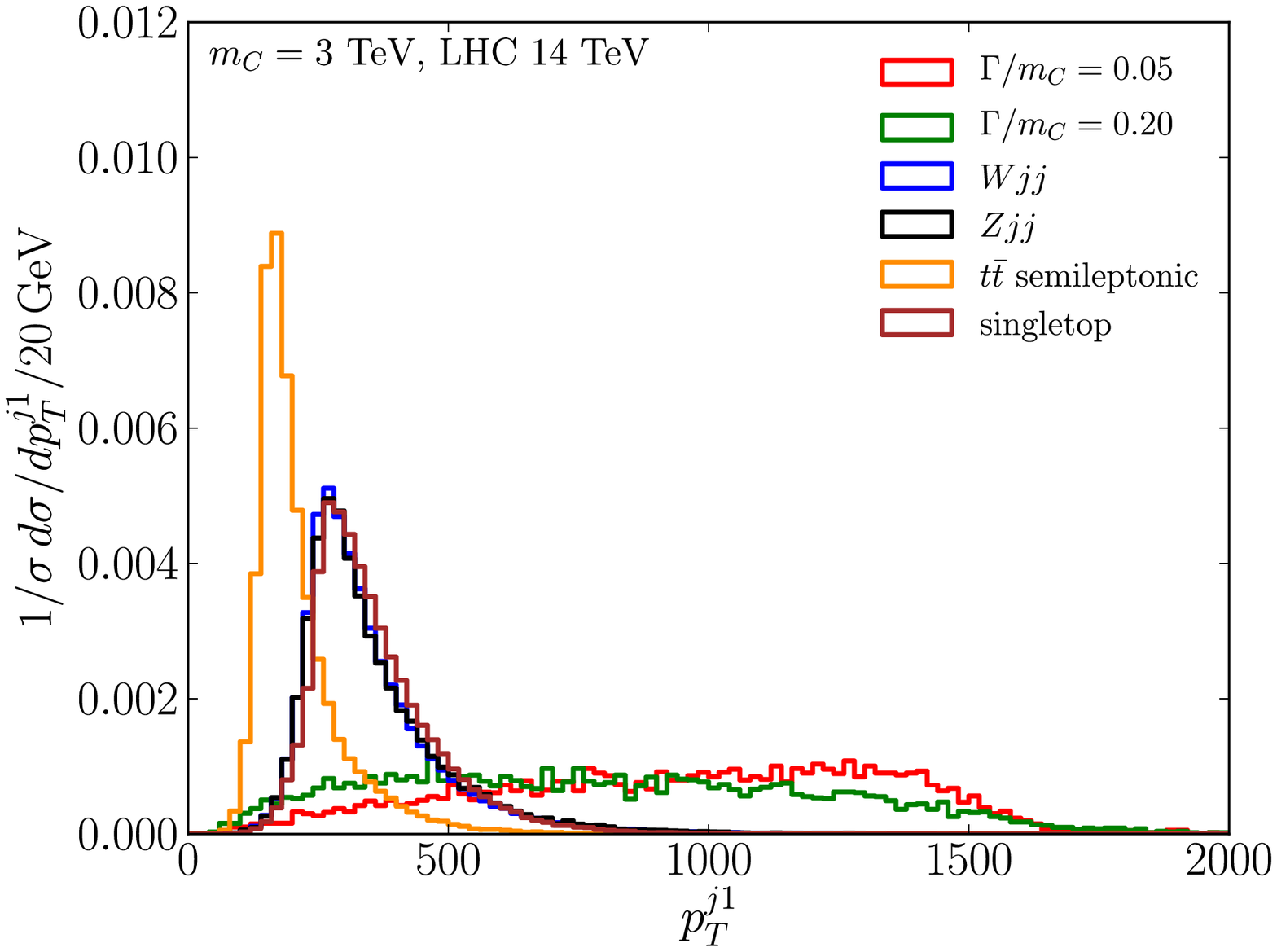}
\includegraphics[width=0.495\textwidth,clip=true]{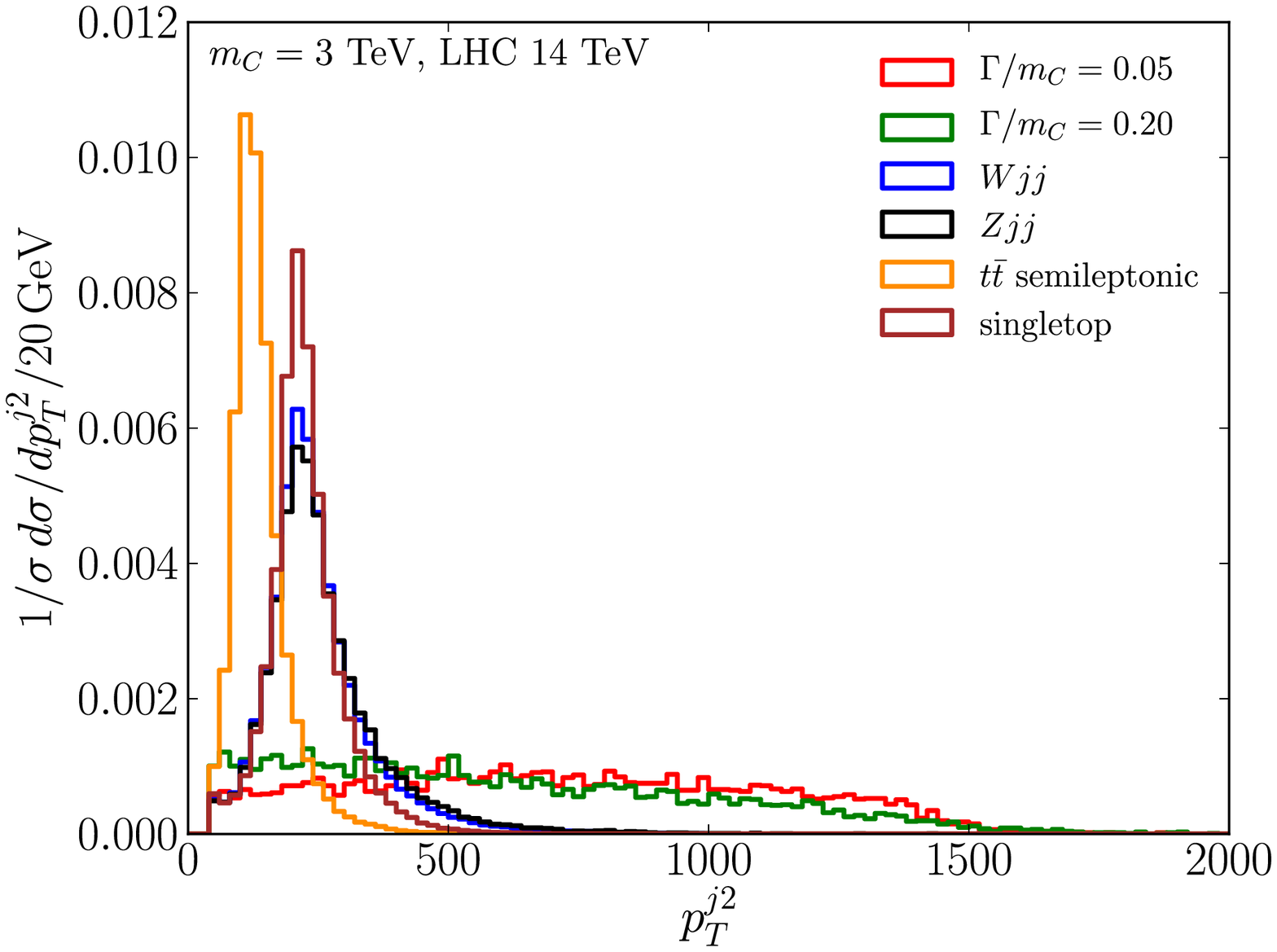}}
\caption{ (a) Top left: transverse momentum of the hardest jet ($p_T^{j_1}$) for a color-octet with mass $m_C = 3\ \tev$ and two different widths $\Gamma_C/m_C = 0.05$ and $0.20$. (b) top right: same as (a) but for the second hardest jet ($p_T^{j_2}$).  The relatively broad and flat transverse momentum distributions of the signal events contrast with the distributions for the backgrounds, which all peak sharply at lower transverse momenta.}  
\label{fig:cwptdist}
\end{figure}

Each event is required to contain at least two jets and exactly one lepton isolated from other leptons and jets, the criteria for which are listed in Eq.~(\ref{eq:basiccuts}). There are also additional minimum requirements on the transverse momentum and rapidity of the jets, leptons and missing energy listed in Eq.~(\ref{eq:basiccuts}). We will refer to these requirements as acceptance cuts. As jets from the decay of a heavy resonance are highly energetic, we identify the two jets with the highest $p_T$ as the jets from the decay of the color-octet. In Fig.~\ref{fig:cwptdist}(a) and (b) we show the transverse momentum distributions for these jets  normalised to unit area for a color-octet with mass $m_C = 3\ \tev$ and two different widths $\Gamma_C/m_C = 0.05$ and $0.20$. It is clear from the distributions that the $p_T$ of the jets is a good discriminant to separate signal from the background. The signal distributions are very broad and the background distributions are sharply peaked at lower transverse momentum. In addition, since associated production would be studied after the existence and mass of the color-octet have been established by dijet studies, we can use transverse momentum cuts that are optimized based on the mass of the color-octet state.

The efficiencies of the acceptance cuts and the optimized transverse momentum cuts for various signal and background samples (optimized to give 3$\sigma$ significance) are shown in Table~\ref{tab:cweff} for different mass values of the color-octet at 14 TeV c.m. energy. The couplings and widths for the mass points shown are the same as in Table~\ref{tab:xsecs}. We can achieve very good separation of signal from background by just using these simple $p_T$ cuts as seen from Table~\ref{tab:cweff}. We find that the commonly used kinematic variable $H_T$, which is the sum of the transverse momenta of final state particles, is dominated by the two hardest jets and there is no significant gain in efficiency by further optimizing on this variable. Similarly, we have investigated the possibility of using the reconstruction of the mass peak as another useful discriminant; as discussed in Sec.\ref{sec:sensitivity}, we found that its likely utility is limited.

\begin{table}[!h] 
\begin{tabular}{|c| c | c  |  c  |  c |}
\hline
\hline
\multicolumn{5}{|c|}{Efficiencies (in \%)} \\
\hline
\hline
Cuts for  & Acceptance cuts & \multirow{2}{*} {$p_T^{j_1} > 825\  \gev$} &\multirow{2}{*} {$p_T^{j_2} > 775\  \gev$} & \multirow{2}{*}{Overall}	\\
$m_C = 2.5\  \tev$ & see Eq.~(\ref{eq:basiccuts}) &  &  & 	\\
\hline
$\epsilon_{2.5}$ & 54 & 53 & 63 & 18 \\
$\epsilon_{Wjj}$ & 54 & 1.3 & 36 & 0.25 \\
$\epsilon_{t\bar{t}}$ & 58 & 8.9$\times 10^{-2}$ & 6.1 & 3.1$\times 10^{-3}$ \\
$\epsilon_{Zjj}$	 & 13 & 1.5 & 49 & 9.2$\times 10^{-2}$ \\
$\epsilon_{t}$ & 56 & 5.0$\times 10^{-1}$ & 6.1 & 1.7$\times 10^{-2}$ \\
\hline
\hline
Cuts for  & Acceptance cuts & \multirow{2}{*} {$p_T^{j_1} > 1000\  \gev$} &\multirow{2}{*} {$p_T^{j_2} > 950\  \gev$} & \multirow{2}{*}{Overall}	\\
$m_C = 3.0\  \tev$ & see Eq.~(\ref{eq:basiccuts}) &  &  & 	\\
\hline
$\epsilon_{3.0}$ & 54 & 41 & 60 & 13 \\
$\epsilon_{Wjj}$ & 54 & 4.6$\times 10^{-1}$ & 35 & 8.8$\times 10^{-2}$ \\
$\epsilon_{t\bar{t}}$ & 58 & 1.9$\times 10^{-2}$ & 7.5 & 7.9$\times 10^{-4}$ \\
$\epsilon_{Zjj}$	 & 13 & 5.5$\times 10^{-1}$ & 48 & 3.3$\times 10^{-2}$ \\
$\epsilon_{t}$ & 56 & 1.1$\times 10^{-1}$ & 4.1 & 2.5$\times 10^{-3}$ \\
\hline
\hline
Cuts for  & Acceptance cuts & \multirow{2}{*} {$p_T^{j_1} > 1100\  \gev$} &\multirow{2}{*} {$p_T^{j_2} > 1050\  \gev$} & \multirow{2}{*}{Overall}	\\
$m_C = 3.5\  \tev$ & see Eq.~(\ref{eq:basiccuts}) &  &  & 	\\
\hline
$\epsilon_{3.5}$ & 51 & 30 & 62 & 9.8 \\
$\epsilon_{Wjj}$ & 54 & 2.7$\times 10^{-1}$ & 35 & 5.2$\times 10^{-2}$ \\
$\epsilon_{t\bar{t}}$ & 58 & 9.0$\times 10^{-3}$ & 1.9 & 9.5$\times 10^{-5}$ \\
$\epsilon_{Zjj}$	 & 13 & 3.3$\times 10^{-1}$ & 48 & 2.0$\times 10^{-2}$ \\
$\epsilon_{t}$ & 56 & 5.0$\times 10^{-2}$ & 4.5 & 1.3$\times 10^{-3}$ \\
\hline
\hline
Cuts for  & Acceptance cuts & \multirow{2}{*} {$p_T^{j_1} > 1200\  \gev$} &\multirow{2}{*} {$p_T^{j_2} > 1150\  \gev$} & \multirow{2}{*}{Overall}	\\
$m_C = 4.0\  \tev$ & see Eq.~(\ref{eq:basiccuts}) &  &  & 	\\
\hline
$\epsilon_{4.0}$ & 49 & 21 & 61 & 6.5 \\
$\epsilon_{Wjj}$ & 54 & 1.6$\times 10^{-1}$ & 34 & 3.0$\times 10^{-2}$ \\
$\epsilon_{t\bar{t}}$ & 58 & 6.8$\times 10^{-3}$ & 0 & 0 \\
$\epsilon_{Zjj}$	 & 13 & 2.1$\times 10^{-1}$ & 42 & 1.1$\times 10^{-2}$ \\
$\epsilon_{t}$ & 56 & 2.4$\times 10^{-2}$ & 1.8 & 2.5$\times 10^{-4}$ \\
\hline
\hline
Cuts for  & Acceptance cuts & \multirow{2}{*} {$p_T^{j_1} > 1350\  \gev$} &\multirow{2}{*} {$p_T^{j_2} > 1300\  \gev$} & \multirow{2}{*}{Overall}	\\
$m_C = 4.5\  \tev$ & see Eq.~(\ref{eq:basiccuts}) &  &  & 	\\
\hline
$\epsilon_{4.5}$ & 50 & 16 & 59 & 4.5 \\
$\epsilon_{Wjj}$ & 54 & 8.0$\times 10^{-2}$ & 31 & 1.4$\times 10^{-2}$ \\
$\epsilon_{t\bar{t}}$ & 58 & 7.1$\times 10^{-4}$ & 0 & 0 \\
$\epsilon_{Zjj}$	 & 13 & 9.0$\times 10^{-2}$ & 35 & 4.0$\times 10^{-3}$ \\
$\epsilon_{t}$ & 56 & 5.4$\times 10^{-3}$ & 8.3 & 2.5$\times 10^{-4}$ \\
\hline
\end{tabular}
\caption{Selection efficiencies (in percent) for signal ($\epsilon_{m_C}$) and background ($\epsilon_{BG}$) samples for the case of the $CW$ channel with leptonic decays of $W$ into electrons and muons. The couplings and widths for the mass points shown are the same as those in Table~\ref{tab:xsecs}. }
\label{tab:cweff}
\end{table}

Finally we mention some details about the signal simulation sample. We study different color-octet masses from $2.5\ \tev$ to $4.5\ \tev$. While we show only a few points for illustration in Table~\ref{tab:cweff}, we sample a wide range of couplings $g^q_{L,R}, g^t_{L,R} $ and decay widths ($\Gamma_C$) ranging from very narrow widths $\sim$ 2.5\% to upwards of 40\% for each mass point. We estimate the significance for each point in parameter space by $s/\sqrt{b}$, where $s\ (b)$ are the number of signal (background) events. The results of this analysis are presented in Sec.\ref{sec:sensitivity}.

\subsection{Neutral Current Channel with $Z$-Decays to Charged Leptons }
\label{sec:collph-cz}

In this section we study the case where the color-octet is produced in association with a $Z$ boson and investigate the prospects of this channel at the LHC with $\sqrt{s} = 14\  \tev$. The color-octet decays to two jets and the $Z$  boson decays leptonically $Z\to \ell \ell\ (\ell=e,\mu)$ giving us a final state $jj\ell\ell$. All the details of the signal simulation are the same as in Sec.\ref{sec:collph-cw}, except that we sample color-octet masses from $2.5\ \tev$ to $4.0\ \tev$ as there is no sensitivity in this channel for higher mass resonances. 

The event is required to contain at least two jets and exactly two leptons isolated from other leptons and jets in the event, the criteria for which are listed in Eq.~(\ref{eq:basiccuts}). We require that the two isolated leptons reconstruct a $Z$ boson with the condition that $m_Z - 20\ {\gev} < m_{\ell\ell} < m_Z + 20\ {\gev}$. There are also additional minimum requirements on the transverse momentum and rapidity of the jets and leptons listed in Eq.~(\ref{eq:basiccuts}).  We will refer to these combined requirements as acceptance cuts. The $p_T$ of the two hardest jets is a good discriminant to separate signal from the background as in the case of the $CW$ channel. The efficiencies of the acceptance cuts and the optimized transverse momentum cuts for various signal and background samples (that give 3$\sigma$ significance) are shown in Table~\ref{tab:czeff} for different mass values of the color-octet at 14 TeV c.m. energy. The couplings and widths for the mass points shown are the same as those in Table~\ref{tab:xsecs}. Again, we sample a much broader range of couplings and widths and show only a few points for illustration. 
\begin{table}[!h]
\begin{tabular}{|c| c | c  |  c  |  c |}
\hline
\hline
\multicolumn{5}{|c|}{Efficiencies (in \%)} \\
\hline
\hline
Cuts for  & Acceptance cuts & \multirow{2}{*} {$p_T^{j_1} > 850\  \gev$} &\multirow{2}{*} {$p_T^{j_2} > 800\  \gev$} & \multirow{2}{*}{Overall}	\\
$m_C = 2.5\  \tev$ & see Eq.~(\ref{eq:basiccuts}) &  &  & 	\\
\hline
$\epsilon_{2.5}$ & 34 & 47 & 68 & 11 \\
$\epsilon_{Zjj}$	 & 42 & 8.4$\times 10^{-1}$ & 43 & 1.5$\times 10^{-1}$ \\
\hline
\hline
Cuts for  & Acceptance cuts & \multirow{2}{*} {$p_T^{j_1} > 975\  \gev$} &\multirow{2}{*} {$p_T^{j_2} > 925\  \gev$} & \multirow{2}{*}{Overall}	\\
$m_C = 3.0\  \tev$ & see Eq.~(\ref{eq:basiccuts}) &  &  & 	\\
\hline
$\epsilon_{3.0}$ & 34 & 38 & 69 & 8.9 \\
$\epsilon_{Zjj}$	 & 42 & 3.9$\times 10^{-1}$ & 45 & 7.3$\times 10^{-2}$ \\
\hline
\hline
Cuts for  & Acceptance cuts & \multirow{2}{*} {$p_T^{j_1} > 1050\  \gev$} &\multirow{2}{*} {$p_T^{j_2} > 1000\  \gev$} & \multirow{2}{*}{Overall}	\\
$m_C = 3.5\  \tev$ & see Eq.~(\ref{eq:basiccuts}) &  &  & 	\\
\hline
$\epsilon_{3.5}$ & 33 & 27 & 70 & 6.2 \\
$\epsilon_{Zjj}$	 & 42 & 2.5$\times 10^{-1}$ & 44 & 4.6$\times 10^{-2}$ \\
\hline
\hline
Cuts for  & Acceptance cuts & \multirow{2}{*} {$p_T^{j_1} > 1200\  \gev$} &\multirow{2}{*} {$p_T^{j_2} > 1150\  \gev$} & \multirow{2}{*}{Overall}	\\
$m_C = 4.0\  \tev$ & see Eq.~(\ref{eq:basiccuts}) &  &  & 	\\
\hline
$\epsilon_{4.0}$ & 34 & 17 & 74 & 4.4 \\
$\epsilon_{Zjj}$	 & 42 & 1.1$\times 10^{-1}$ & 49 & 2.2$\times 10^{-2}$ \\
\hline
\end{tabular}
\caption{Selection efficiencies (in percent) for signal ($\epsilon_{m_C}$) and background ($\epsilon_{BG}$) samples for the case of the $CZ$ channel with leptonic decays of $Z$ into electron and muon pairs. The couplings and widths for the mass points shown are the same as those in Table~\ref{tab:xsecs}. }
\label{tab:czeff}
\end{table}

It is evident from Table~\ref{tab:czeff} that we can achieve very good separation of signal from background for the $CZ$ channel as well by using just simple $p_T$ cuts as in the case of $CW$ channel. Again, we investigated the possible improvements one could make by using the invariant mass of the color-octet resonance as a discriminant. The situation for the case of the associated production of the $Z$ boson with leptonic decays is slightly better, as full information for reconstructing the final state is readily available from the reconstructed leptons. However, as detailed in Sec.\ref{sec:sensitivity}, we determined that it would not add appreciably to the present analysis. Next, we discuss the results of our analysis and the sensitivity at the LHC.

\section{Sensitivity at the LHC}
\label{sec:sensitivity}

In this section we will describe the sensitivity of the associated production channel at the LHC, the current constraints from direct searches and future sensitivity at the LHC to provide a broad view of the reach of the LHC in determining the chiral structure of the couplings of a color-octet resonance. We have simulated signal and background events for two different scenarios at the LHC with $\sqrt{s} = 14\ \tev$: an early run with an integrated luminosity of 10 $\invfb$ and a longer run with 100 $\invfb$ integrated luminosity. We estimate the number of signal and background events after optimizing the transverse momentum cuts as described in Sec.\ref{sec:collph-cw} and \ref{sec:collph-cz} and the statistical significance is calculated as $s/\sqrt{b}$ where $s\ (b)$ is the number of signal (background) events. 

\begin{figure}[tb]
\includegraphics[width=0.495\textwidth,clip=true]{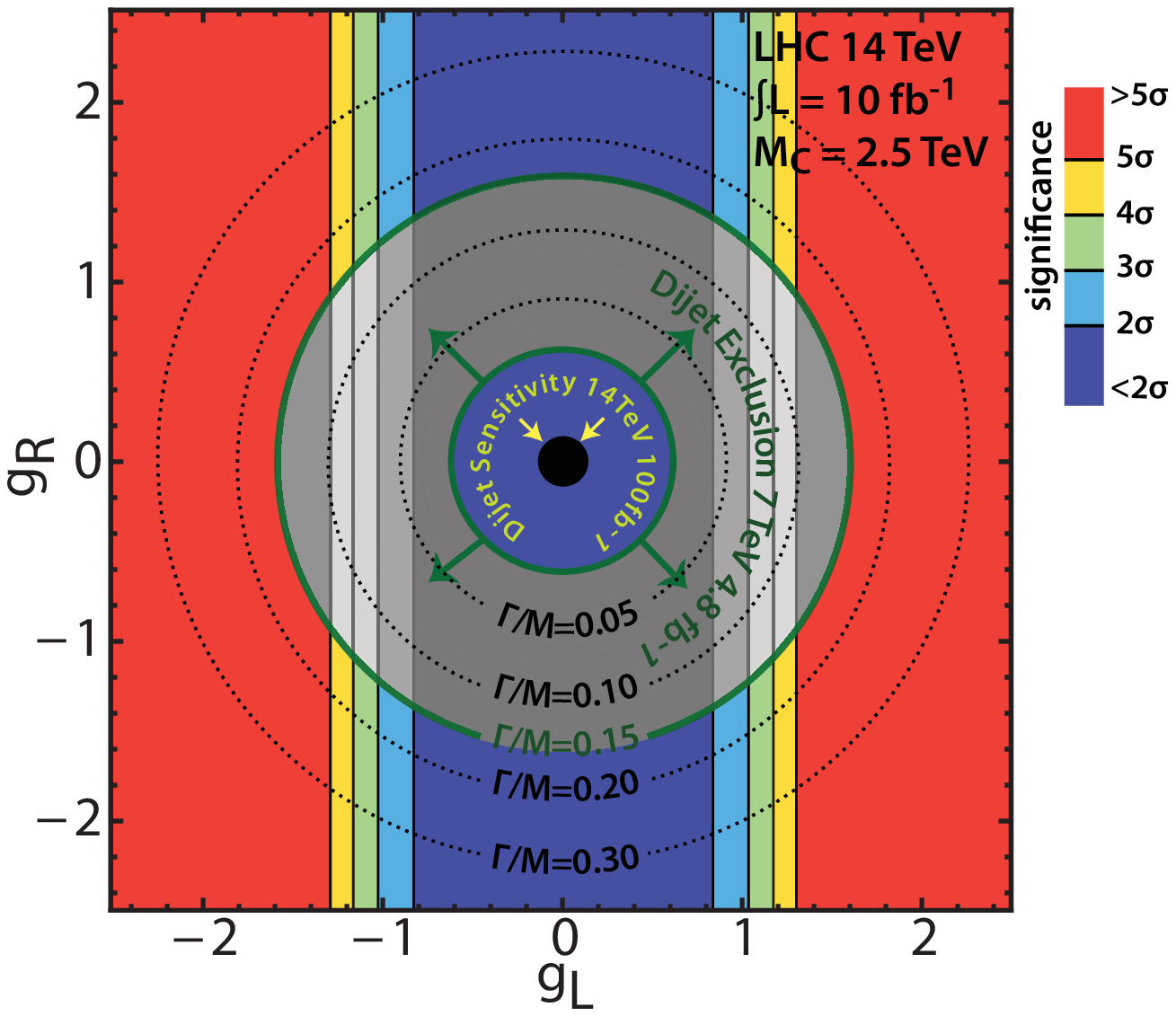}
\includegraphics[width=0.495\textwidth,clip=true]{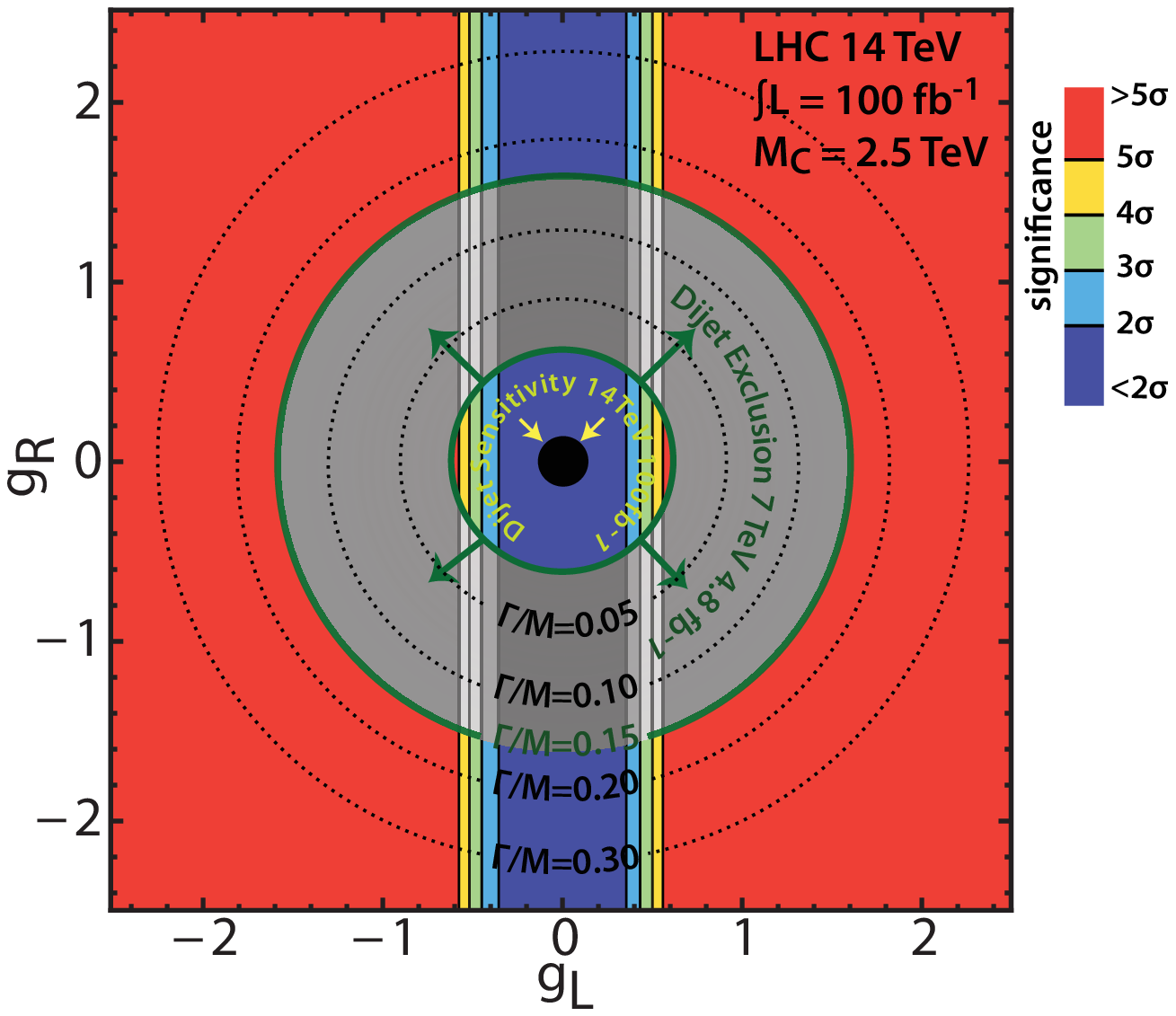}
\includegraphics[width=0.495\textwidth,clip=true]{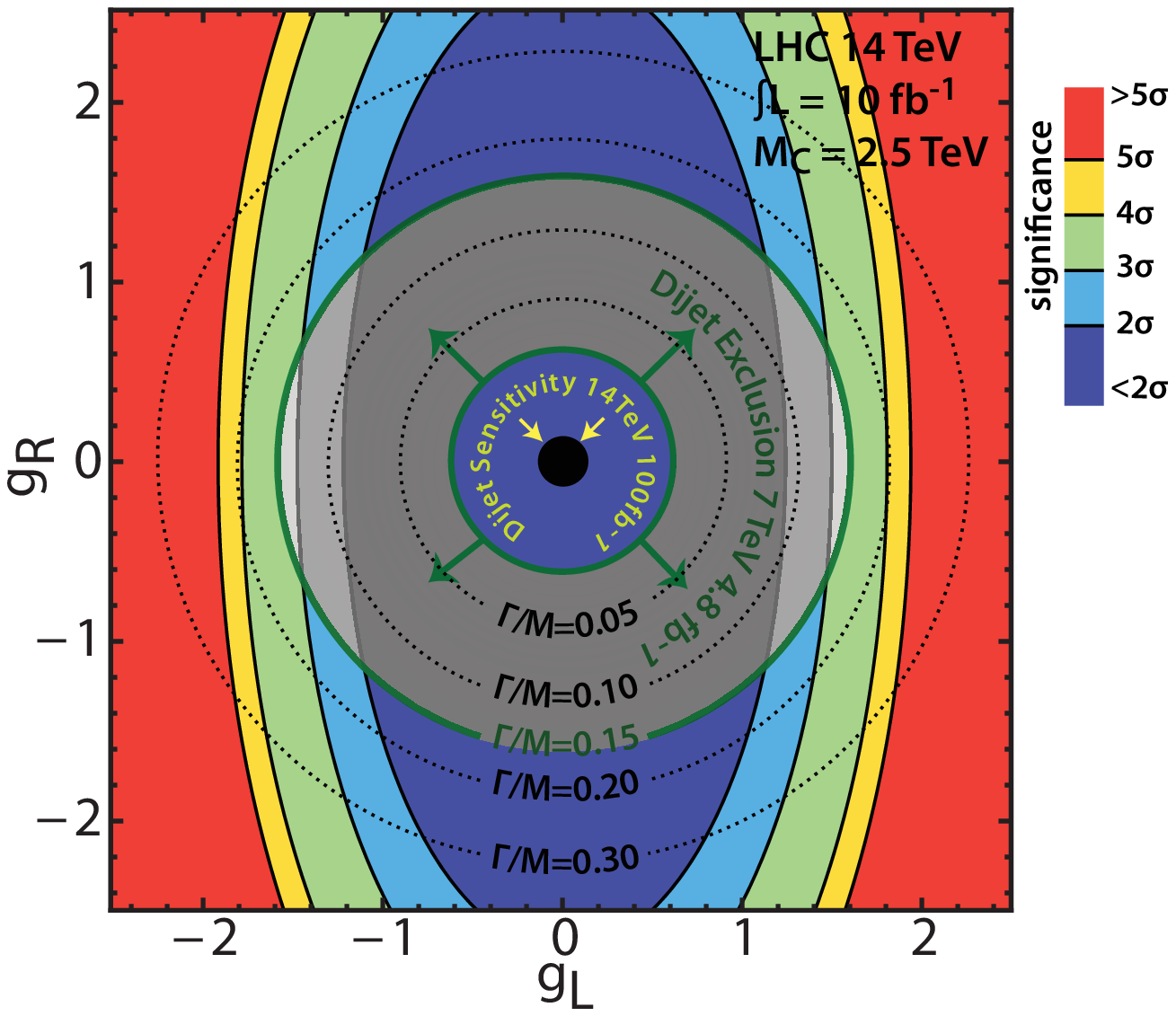}
\includegraphics[width=0.495\textwidth,clip=true]{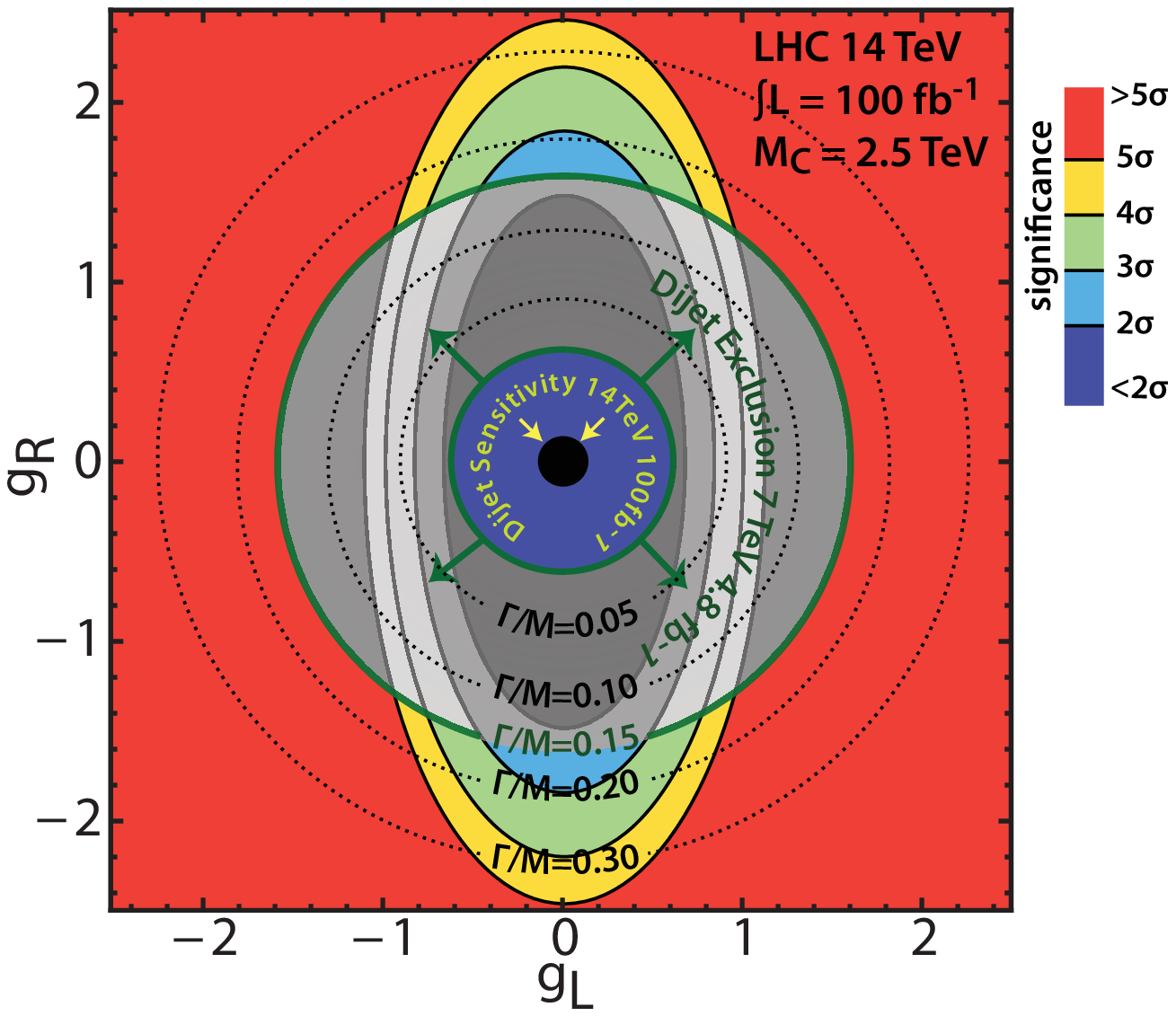}
\caption{ 
(a) Top left: sensitivity plot for a color-octet of mass $m_C = 2.5\ \tev$ produced in association with a $W$ gauge boson, in the plane of the couplings $g_L^q, g_R^q$ at the LHC with 10 $\invfb$ of data and $\sqrt{s} =  14\ \tev$; (b) top right: same
  as (a) but for 100 $\invfb$ of data;
  (c) bottom left: same as (a) but for a color-octet produced in association with a $Z$ boson; (d) bottom right: same as (c) but for
  100 $\invfb$ of data. The various color bands show the regions with varying significance from $2\sigma$ to $>5\sigma$. The inner solid green circle is the limit from current direct searches for narrow dijet resonances at the LHC \cite{Aad:2011fq, Aad:2010ae, Chatrchyan:2011ns, Khachatryan:2010jd, atlas:2012-038}. The outer green circle corresponds to a contour with $\Gamma_C/m_C = 0.15$. The region between the two circles represented as a faded grey region is excluded for narrow resonances while the region beyond the outer green circle is allowed for broad resonances. The small black region in the center lies beyond the projected dijet sensitivity at the LHC with 100 $\invfb$ of data \cite{cmsnote}. The black dotted contours indicate the combinations of couplings that give rise to varying widths $\Gamma_C/m_C = 0.05,\  0.10,\  0.20$ and $0.30$. 
 }  
\label{fig:cwcz2.5tev}
\end{figure}

\begin{figure}[tb]
\includegraphics[width=0.495\textwidth,clip=true]{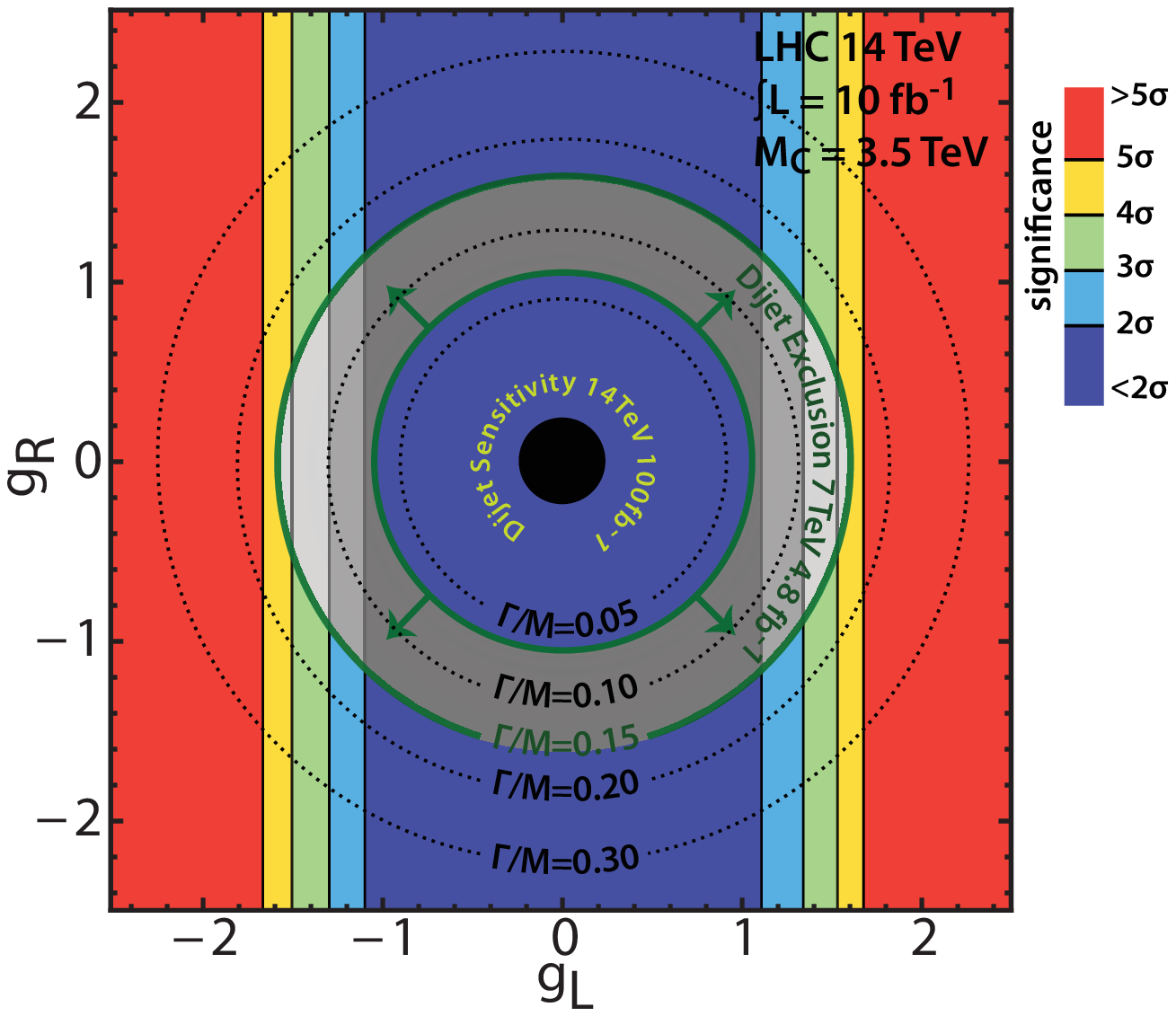}
\includegraphics[width=0.495\textwidth,clip=true]{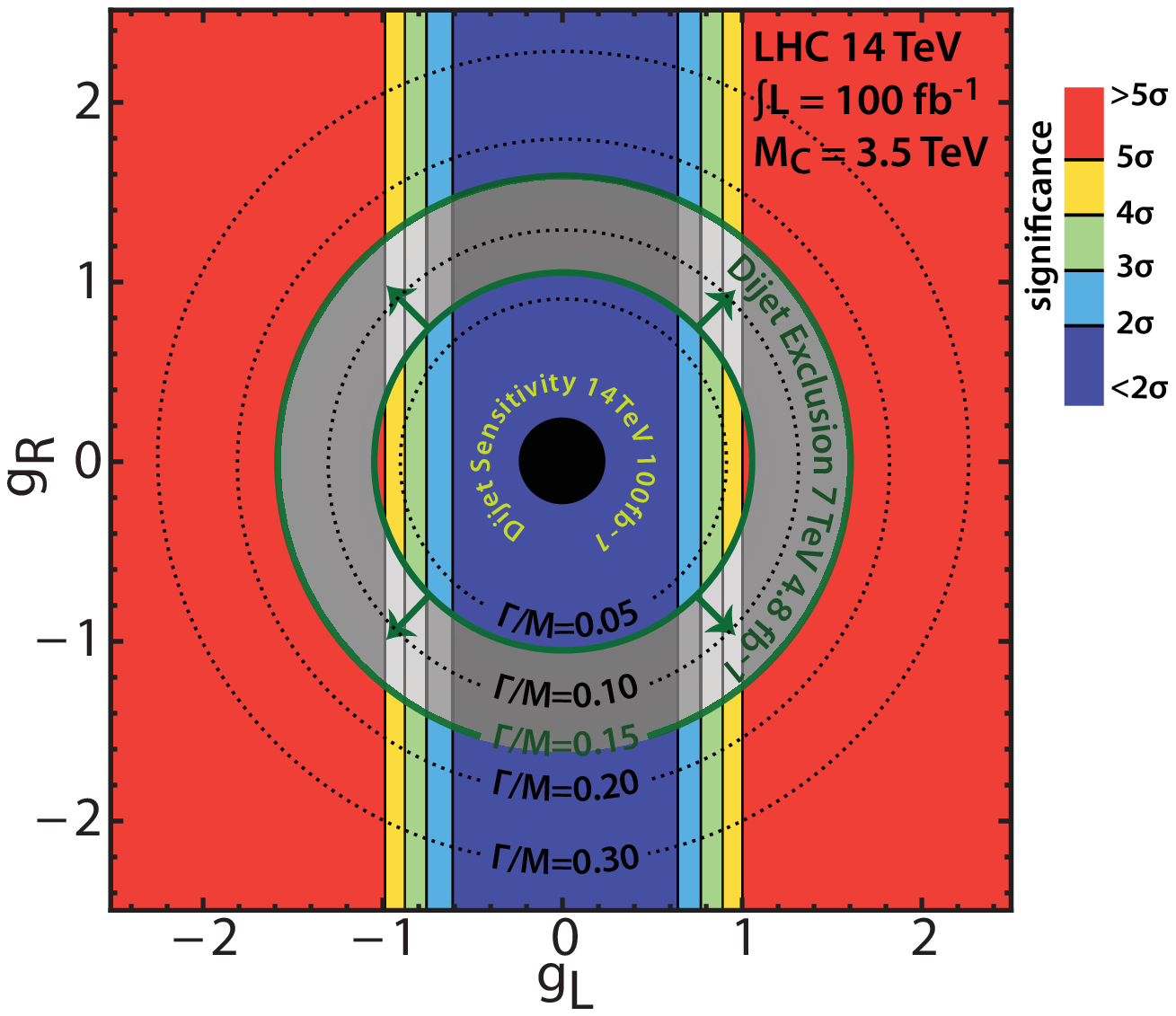}
\includegraphics[width=0.495\textwidth,clip=true]{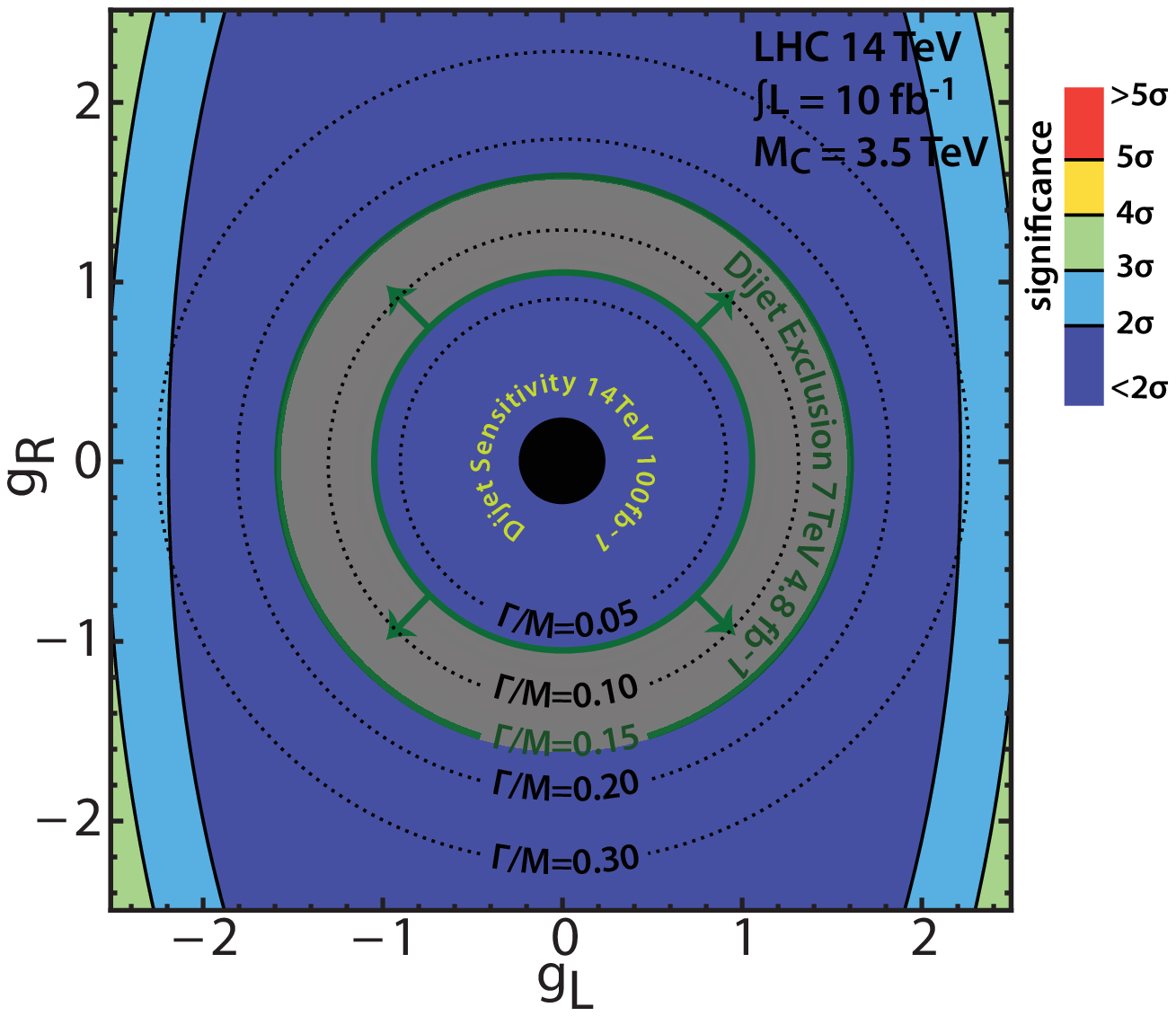}
\includegraphics[width=0.495\textwidth,clip=true]{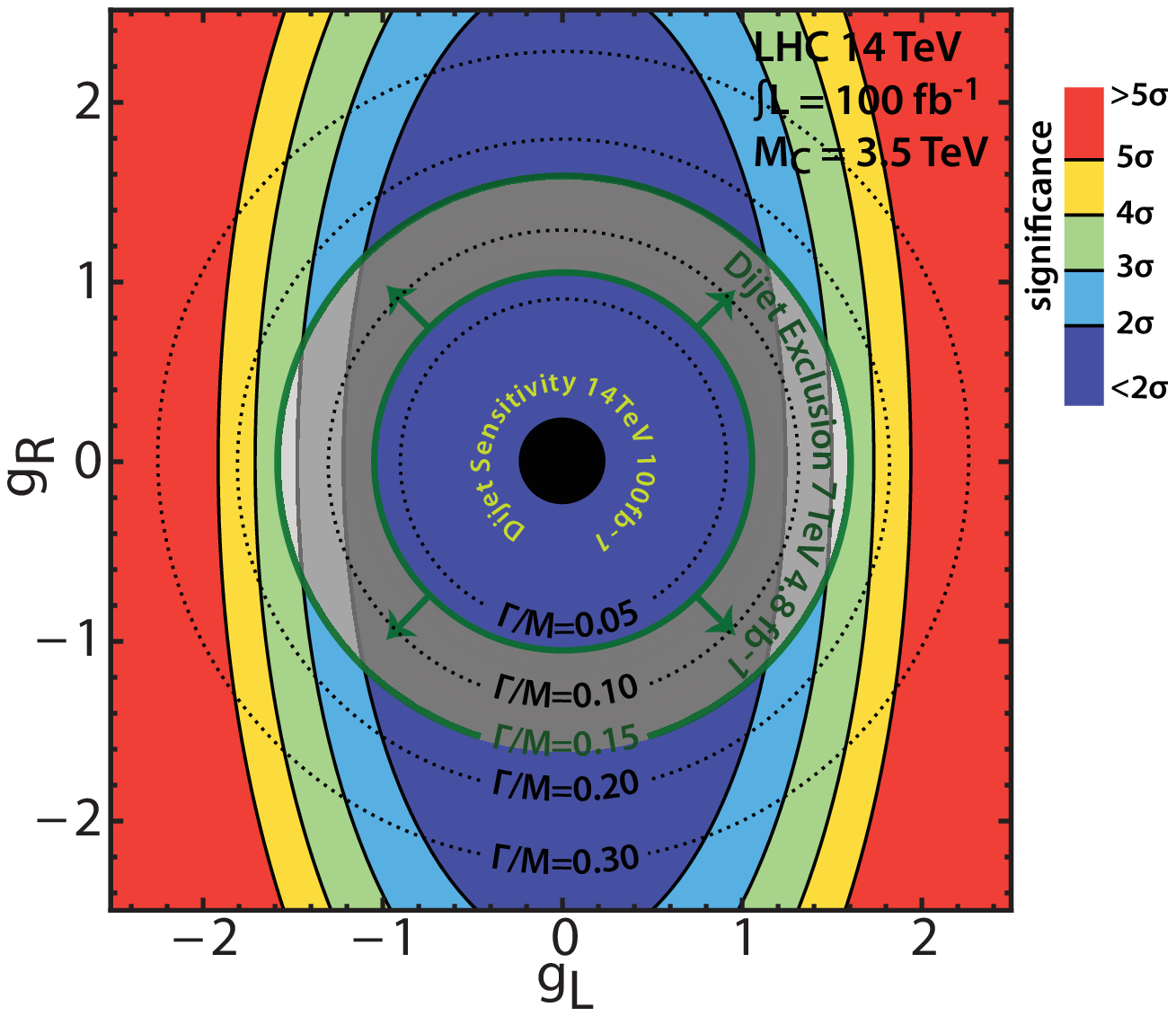}
\caption{ 
Same as Fig.~\ref{fig:cwcz2.5tev} but for $m_C = 3.5\ \tev$.
 }  
\label{fig:cwcz3.5tev}
\end{figure}

\begin{figure}[tb]
\includegraphics[width=0.495\textwidth,clip=true]{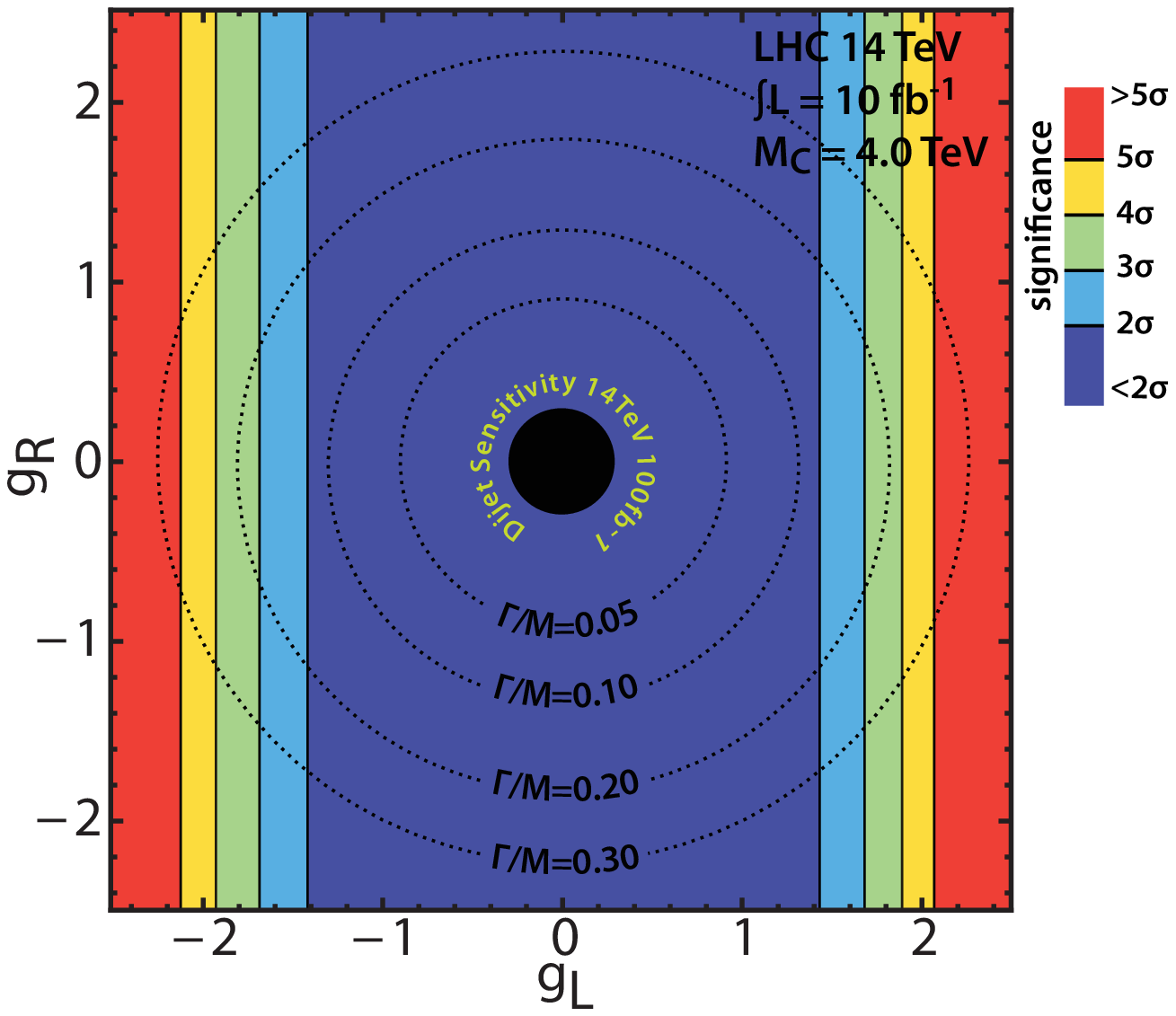}
\includegraphics[width=0.495\textwidth,clip=true]{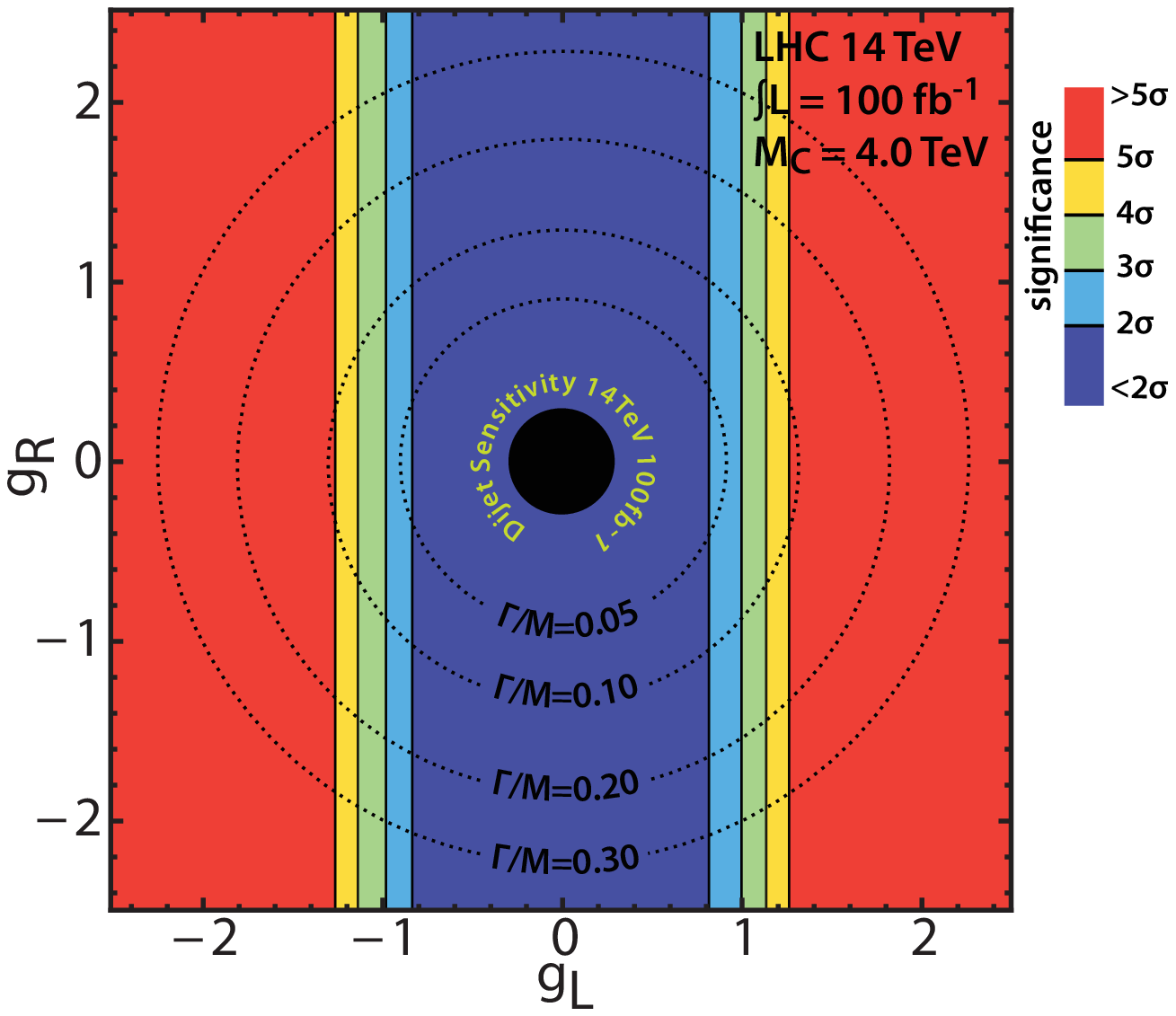}
\includegraphics[width=0.495\textwidth,clip=true]{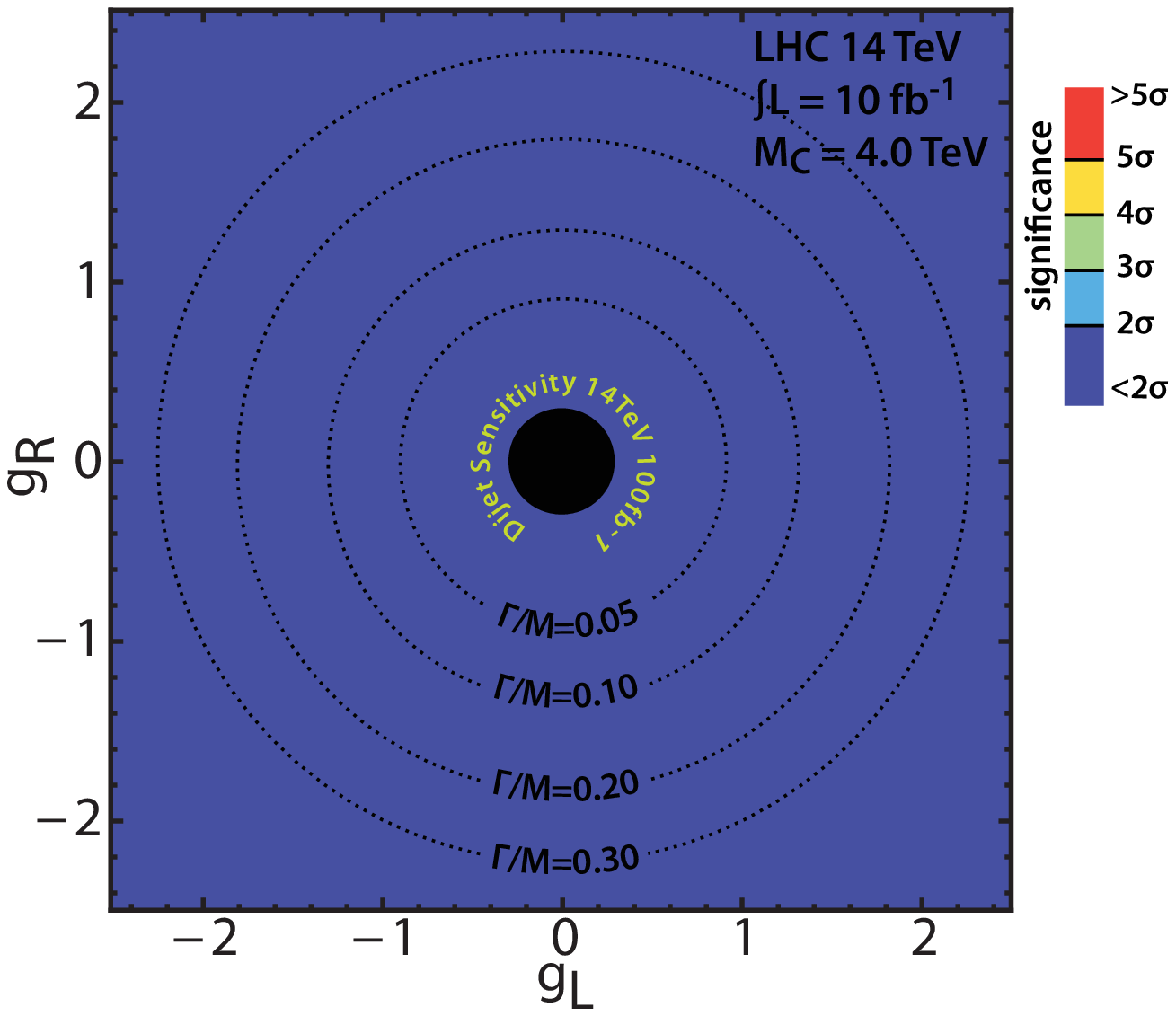}
\includegraphics[width=0.495\textwidth,clip=true]{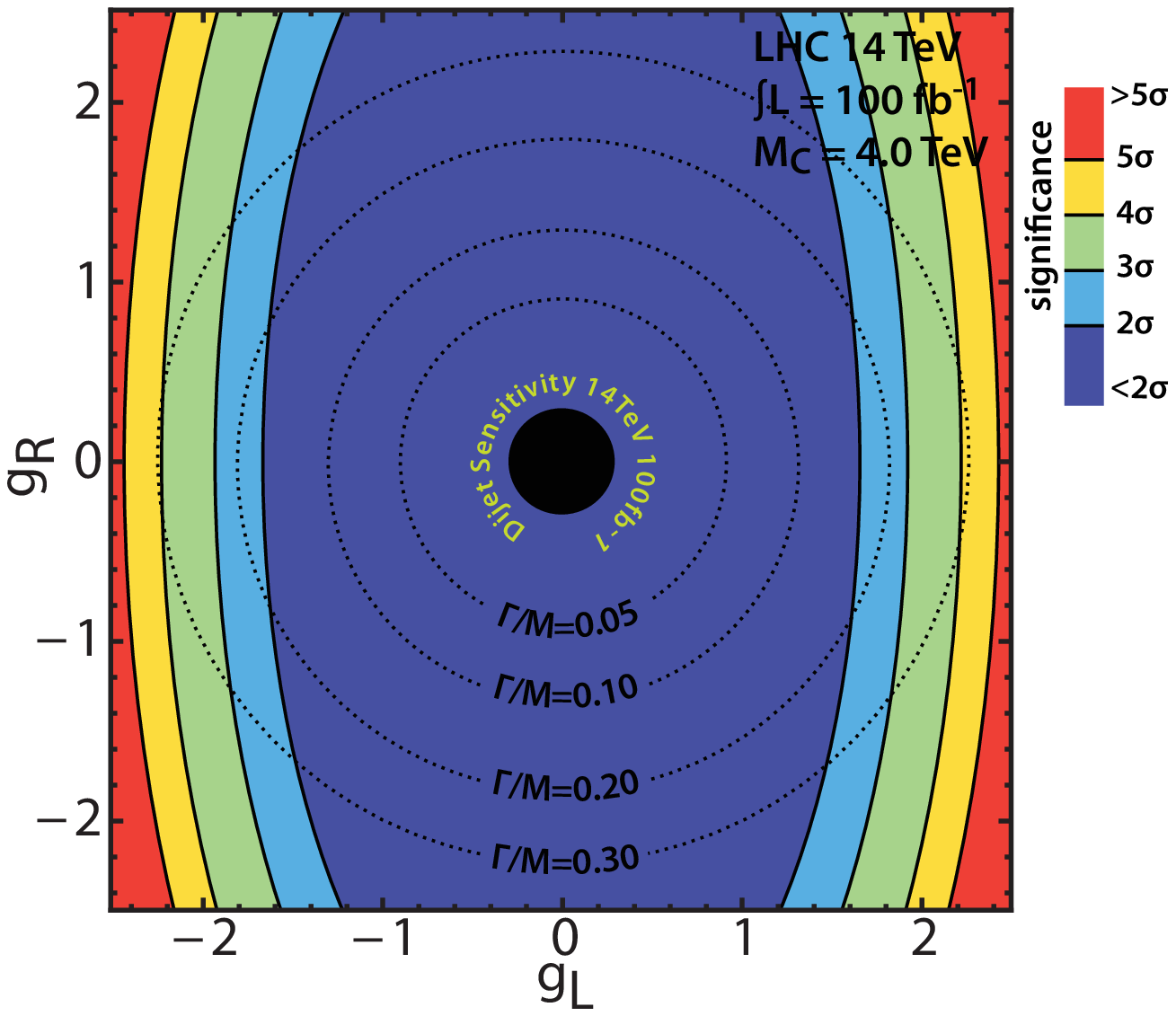}
\caption{ 
Same as Fig.~\ref{fig:cwcz2.5tev} but for $m_C = 4\ \tev$.   
}  
\label{fig:cwcz4tev}
\end{figure}

\begin{figure}[tb]
{\includegraphics[width=0.495\textwidth,clip=true]{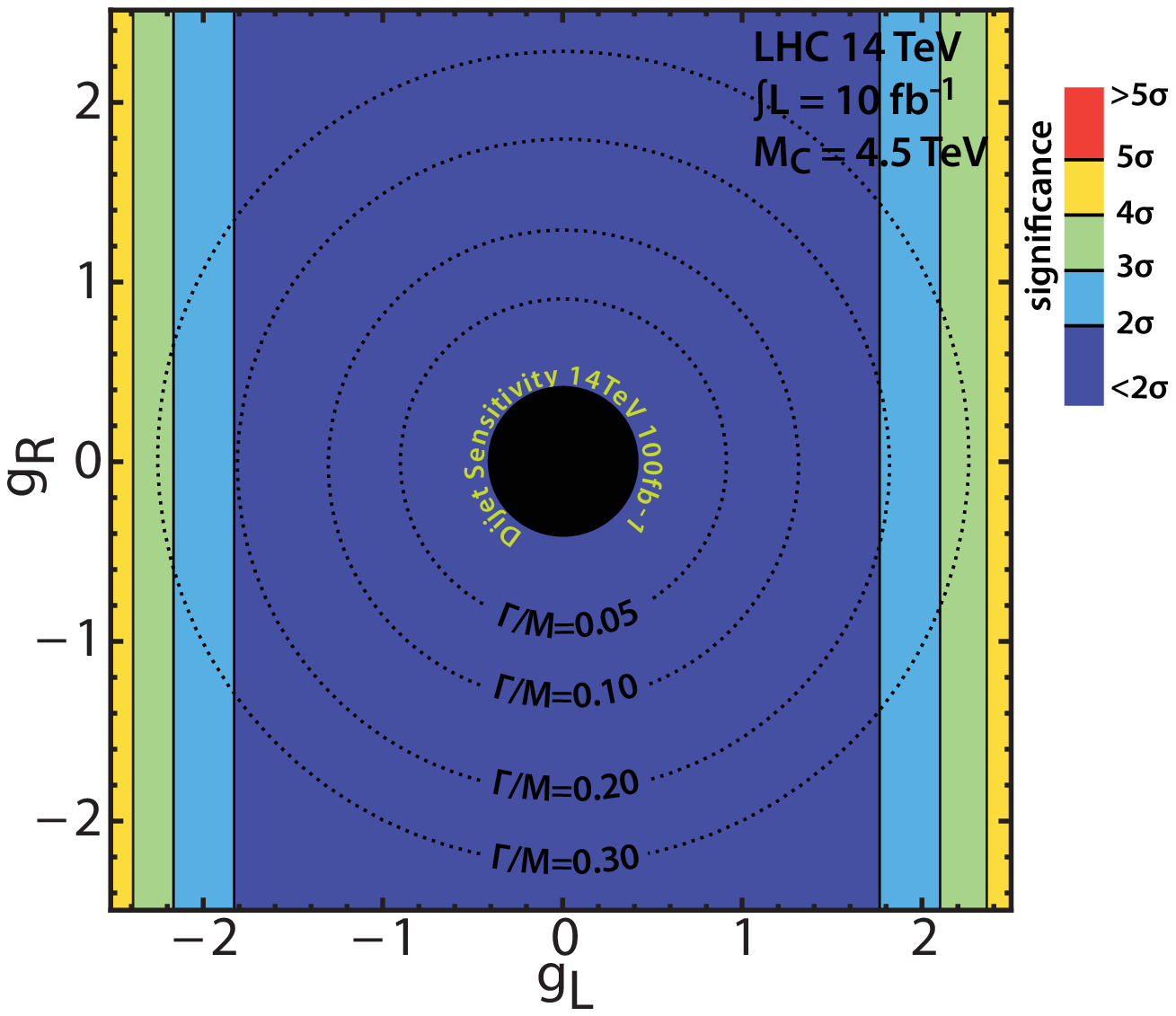}
\includegraphics[width=0.495\textwidth,clip=true]{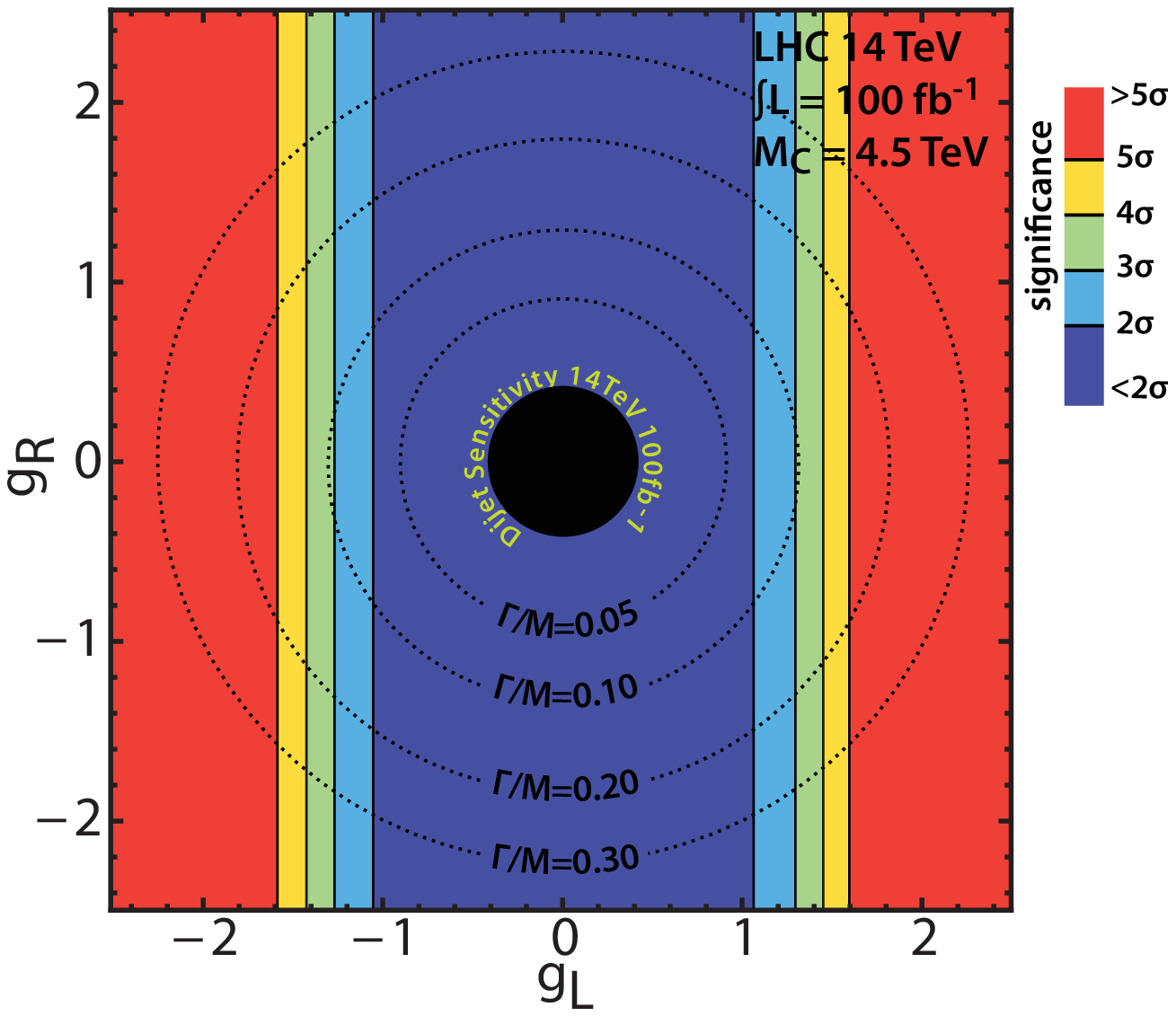}}
\caption{ (a) Left: same as Fig.~\ref{fig:cwcz2.5tev}(a) but for $m_C = 4.5\ \tev$; (b) right: same as (a) but for 100 $\invfb$ of data. The figures for the corresponding $CZ$ channels are not presented as there is not sufficient sensitivity at this mass.}  
\label{fig:cw4.5tev}
\end{figure}

\begin{figure}[tb]
{\includegraphics[width=0.495\textwidth,clip=true]{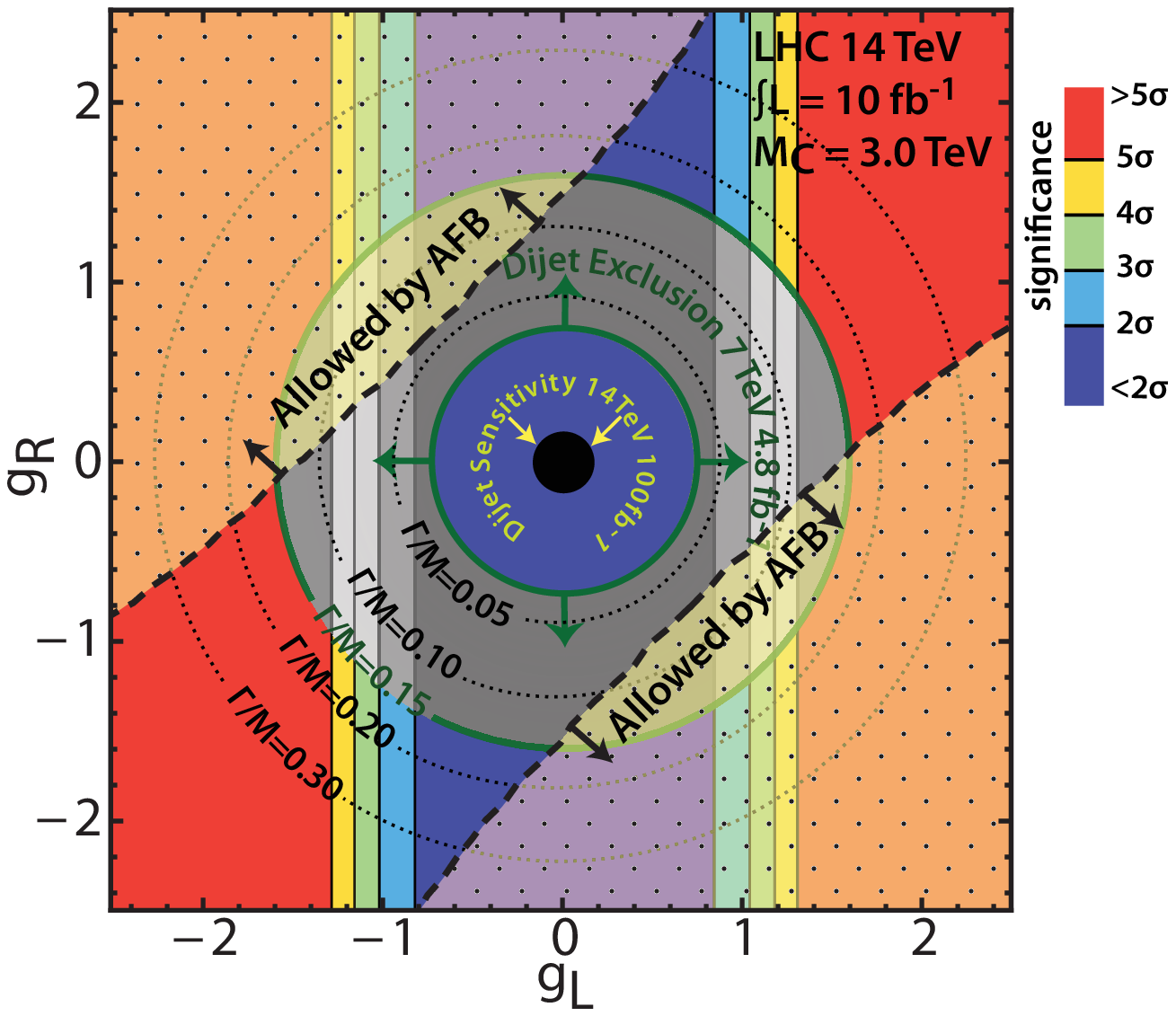}
\includegraphics[width=0.495\textwidth,clip=true]{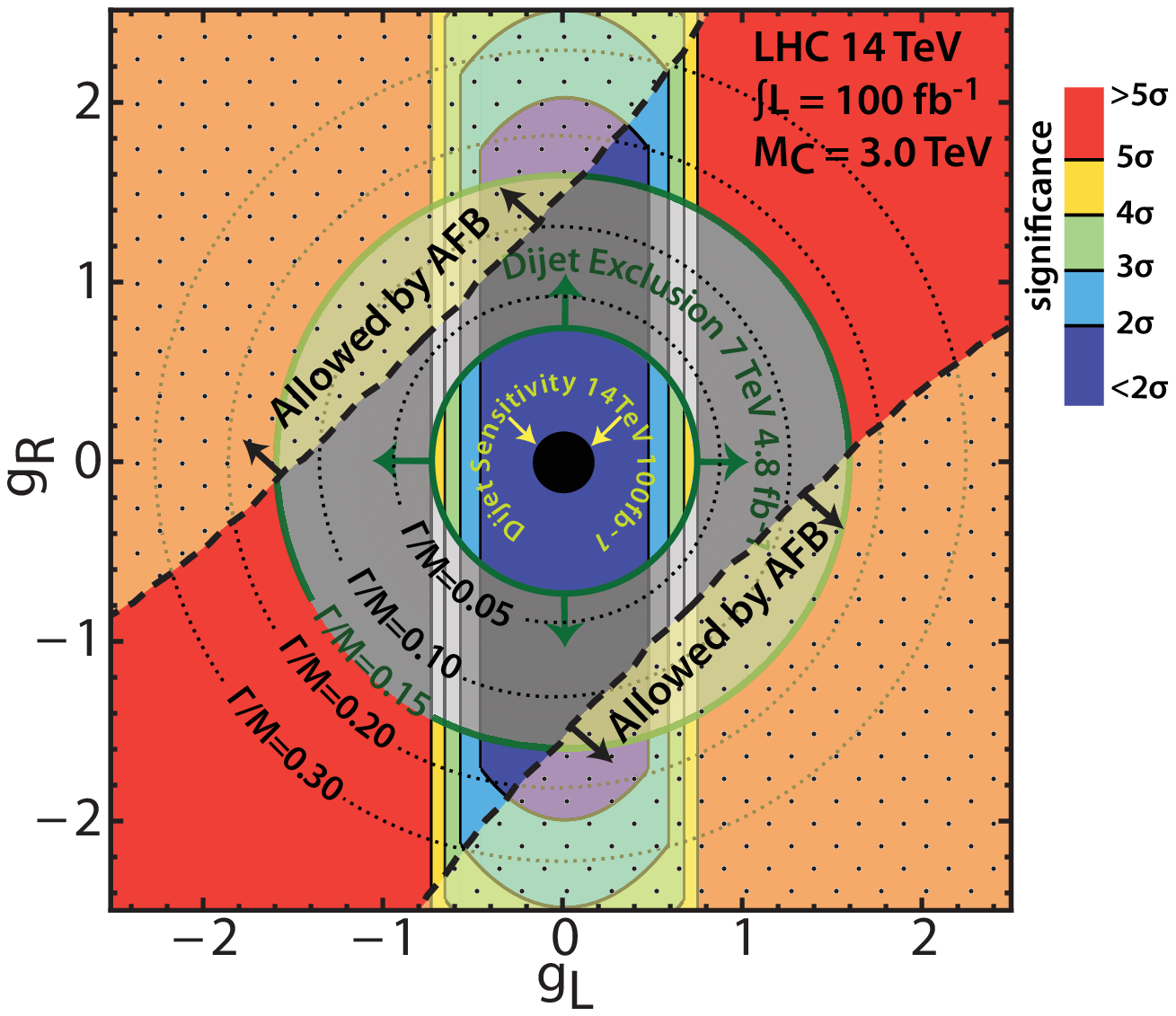}}
\caption{ (a) Left: same as Fig.~\ref{fig:cwcz2.5tev}(a) but for $m_C = 3\ \tev$. Here we have combined the sensitivity in the associated production channel from both $W$ and $Z$ bosons. The couplings allowed by the current measurement of $\afbt$ at the Tevatron \cite{cdf_afb5invfb2010, d0_afb4invfb2010} as interpreted in Ref.~\cite{Rodrigo:2010gm} are shown by the translucent yellow dotted region; (b) right: same
  as (a) but for 100 $\invfb$ of data;
}  
\label{fig:cwcz3tev}
\end{figure}

We present the results of our analysis for the LHC in Figs.~(\ref{fig:cwcz2.5tev}-\ref{fig:cwcz3tev}) in the plane of the couplings $g_L^q, g_R^q$ for different masses, $m_C$, of the color-octet. The sensitivity for the channel with associated production of a $W(Z)$ gauge boson is presented in the upper (lower) panels of Figs.~(\ref{fig:cwcz2.5tev}-\ref{fig:cwcz4tev}), while the left (right) panels are for integrated luminosity of 10 (100) $\invfb$. In Figs.~\ref{fig:cw4.5tev}(a) and (b) we show the sensitivity for a 4.5 TeV color-octet resonance for 10 $\invfb$ and 100 $\invfb$ respectively in the $CW$ channel. The $CZ$ channel has no sensitivity at this mass and we do not show that channel in Fig.~\ref{fig:cw4.5tev}. The different colored bands represent varying significance of signal observation from $2\sigma$ to greater than $5\sigma$. The black dotted curves are contours of constant widths and we show the curves for several values of $\Gamma_C/m_C = 0.05,\  0.10,\  0.20$ and $0.30$. The small black region in the center lies outside the projected dijet sensitivity (i.e. couplings within the black region result in a dijet production rate too small to be observed) at the LHC with 100 $\invfb$ of data \cite{cmsnote}. 

The Tevatron \cite{Aaltonen:2008dn} and the LHC \cite{Aad:2011fq, Aad:2010ae, Chatrchyan:2011ns, Khachatryan:2010jd, atlas:2012-038} have looked for resonances in the dijet spectra, and the most stringent constraints come from the LHC as expected. This data places stronger constraints on low mass resonances and there is essentially no constraint on color-octets with masses above 3.5 TeV. The inner green solid circle is the limit from the current non-observation of narrow resonances in the dijet channel at the LHC \cite{Aad:2011fq, Aad:2010ae, Chatrchyan:2011ns, Khachatryan:2010jd, atlas:2012-038} and the region outside this inner green circle can potentially be excluded.  However there is a caveat here. The analyses of the LHC dijet searches make the assumption of a resonance with a narrow width of order 10\% - 15\%. The authors of Ref.~\cite{Bai:2011ed, Haisch:2011up, Harris:2011bh} argue that in the case where the resonance is not narrow ($\Gamma_C/m_C >15\%$) the constraints from dijet data can be relaxed. We mark the contour of $\Gamma_C/m_C = 0.15$ by the outer solid green circle. The area between the two green circles which is shown as a faded gray region (where the narrow-width approximation is valid) is excluded. The region beyond the outer green circle which corresponds to resonances with broader widths is possibly allowed. For example, in Fig.~\ref{fig:cwcz2.5tev}(a) the dijet constraint would be valid for narrow resonances (up to the outer green circle labeled $\Gamma_C/m_C = 0.15$) and would not be applicable to the regions outside this curve. As the LHC accumulates more data, the simple dijet analyses would remain sensitive only to the region inside the $\Gamma_C/m_C = 0.15$ curve. Of course a different analysis of dijet data without the narrow width assumption could ultimately be sensitive to the whole region. The region corresponding to broad resonances (beyond the outer green circle marked $\Gamma_c/m_C = 0.15$) has large couplings and hence will be largely accessible at the LHC. 

The results in Figs.~(\ref{fig:cwcz2.5tev}-\ref{fig:cwcz3tev}) illustrate several features. Firstly, in the channel with associated production of a $W$ boson, there is no sensitivity in the region near $g_L^q = 0$ due to the left-handed couplings of the $W$ boson, and the sensitivity improves as we move away from the $g_L^q = 0$ axis. The channel with the associated production of a $Z$ boson on the other hand is sensitive to both left and right-handed couplings and sensitivity in the region close to $g_L^q = 0$ is non-zero. The smaller production cross section for this channel along with the small leptonic branching fractions for $Z$ decays limit the gain in sensitivity. Nonetheless this channel provides an additional measurement and hence useful information in untangling the couplings.

We present the results for both 10 $\invfb$ and 100 $\invfb$ of data, and as expected, the longer run with more data has better sensitivity and can probe masses up to 4.5 (4.0) TeV in the channel with associated production of a $W(Z)$ gauge boson. We also show the projected limit of sensitivity in the dijet channel for the LHC with 100 $\invfb$ data as the small black circle in the center with small couplings. This can be improved even further, essentially allowing one to get closer to near zero couplings by the following observation. The theoretical analysis that produced the projected sensitivity \cite{cmsnote} selects for a rather narrow region in pseudo-rapidity of the jets leading to a small acceptance. The analyses of actual LHC data for the 7 TeV dijet spectrum at the LHC  \cite{Aad:2011fq, Aad:2010ae, Chatrchyan:2011ns, Khachatryan:2010jd, atlas:2012-038} include larger regions of pseudo-rapidity giving rise to an acceptance larger than the one in Ref.~\cite{cmsnote}. It is safe to assume that the final acceptance will be at least equal to or even better than the current one and the small black region in the center could shrink even further. 

Finally, the dijet reach is much better than that of associated production channel at the LHC, essentially probing near zero couplings. If the LHC were to discover a resonance with such small couplings, one would have to find other novel ways of understanding the chiral structure of couplings as the associated production channel does not have sensitivity in those regions of parameter space. 

Separately, we note that the Tevatron has made a measurement of the top-pair forward-backward asymmetry ($\afbt$) \cite{cdf_afb5invfb2010, d0_afb4invfb2010}. The authors of Ref.~\cite{Rodrigo:2010gm} have translated this measurement into constraints on the couplings of color-octet resonances. We show this additional constraint for the case of a color-octet with mass $m_C = 3\ \tev$ in Figs.~\ref{fig:cwcz3tev}(a) and (b) for 10 $\invfb$ and 100 $\invfb$ respectively. The region consistent with the $\afbt$ measurement is shown in translucent yellow with small dots. Note that for clarity of presentation we have combined the sensitivity from $CW$ and $CZ$ channels in Figs.~\ref{fig:cwcz3tev}(a) and (b). There are studies \cite{Haisch:2011up} that do global fits of data from top pair production including cross section, angular measurement, indirect searches and electroweak constraints under certain assumptions. The interested reader can combine our analysis with that of Ref.~\cite{Haisch:2011up} to see how those other constraints (and assumptions) will affect the reach of the LHC for color-octet resonances.

Next, we discuss some possible means for achieving further improvements to the sensitivity. The reconstruction of the mass peak can be a useful discriminant to separate signal from background provided the mass of the resonance is known from prior measurements. The signal for our process comes from both s and t-channel diagrams as shown in Fig.~\ref{fig:sigdiag}. In the case where the s-channel contribution dominates and the associated $W$ gauge boson is from the initial state, it is straightforward to form the invariant mass of the resonance $m_C = m(j_1, j_2)$ for the signal and backgrounds. As the jets from the decay of the color-octet are very energetic, there is hard radiation from the jets and we found that using the three hardest jets reconstructs the invariant mass better and so it is better to use $m_C = m(j_1, j_2, j_3)$ for the signal and backgrounds. In the case where the associated $W$ gauge boson is produced in the final state with $W\to \ell\nu$ decay, we do not have full information to reconstruct the event. Since we cannot determine which of the two configurations the final state $W$ comes from, one can make the conservative choice of using the transverse mass:
\bea
\nn
m_T^2 = \left(\sqrt{p_{T}^{W2}+m_{W}^2} + p_{T}^{j_1} + p_{T}^{j_2} + p_{T}^{j_3} \right)^2 -
\left(\vec p_{T}^{\ W} + \vec p_{T}^{\ j_1} + \vec p_{T}^{\ j_2} + \vec p_{T}^{\ j_3} \right)^2.
\eea
Our initial investigation found that selecting events based on a transverse mass cut  yielded only modest improvement. Even reconstructing events with a leptonically decaying $W$ using the $W$ rest mass approximation did not help much. However further study may be warranted. 
 
\begin{figure}[tb]
{\includegraphics[width=0.60\textwidth,clip=true]{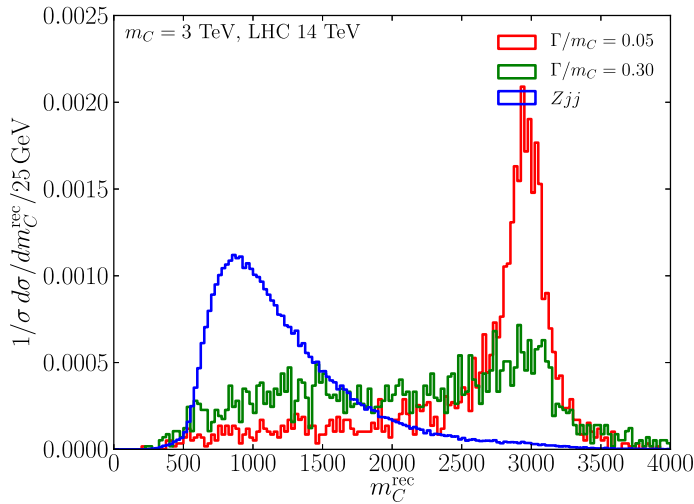}}
\caption{Reconstructed invariant mass ($m_C^{rec}$) for a 3 TeV color-octet resonance with width $\Gamma_C/m_C = 0.05$ and $0.30$ after acceptance cuts. }  
\label{fig:czmcrecdist}
\end{figure}

The situation for the case of the associated production of a $Z$ boson with leptonic decays is slightly better as full information for reconstructing the final state is readily available. In the case where the s-channel contribution dominates, the invariant mass of the resonance can be reconstructed from $m_C = m(j_1, j_2, j_3)$ (initial state Z) or $m_C = m(j_1, j_2, j_3, Z)$ (final state Z). Since we cannot determine which of the two configurations the $Z$ boson comes from, we pick the one closest to the resonance mass and call it $m_C^{rec}$. In Fig.~\ref{fig:czmcrecdist} we show the reconstructed invariant mass ($m_C^{rec}$) for a 3 TeV color-octet resonance with width $\Gamma_C/m_C = 0.05$ and $0.30$. The distribution for $m_C^{rec}$ is normalized to unit area. While it is easy to distinguish the mass peak for the case of a very narrow resonance, it becomes increasingly harder to do so for broader resonances. Furthermore it will be hard to determine the mass of a broad resonance even in the dijet channel. Since we sample a wide range of couplings which lead to varying widths from 2.5\% to greater than 40\% of the mass, using the invariant mass as a discriminant is not useful for the entire range. Also, in the case where the t-channel contribution is not negligible the resonance peak is diluted and the utility of the invariant mass as a discriminant is further reduced. In view of these observations we do not use the invariant mass as a discriminant to separate signal from backgrounds for either the $CW$ or $CZ$ channel but mention it here for completeness as a possible improvement one could make for the appropriate cases (narrow width resonances).

The analysis reported here gives only an estimate of the LHC reach and we discuss some considerations about uncertainties and systematic effects. Our predictions for the signal and background are at the leading order (LO) and no k-factors have been taken into account. Varying the scales by a factor of two around the central value results in a variation of the signal cross sections of order $15-30\%$ and can be used as an estimate of the uncertainty of the leading order (LO) cross sections. For the backgrounds one can use the k-factors where available or they can be normalized to data in regions outside the signal. Multijet events, which are abundant at the LHC, can fake the signal for the $CW$ channel if jets are misreconstructed as leptons and if jet transverse energies are poorly measured, leading to a presence of missing energy in the events. As explained in Ref.~\cite{Atre:2011ae} these effects will be small with optimized experimental methods. Also high luminosity measurements have to contend with the issue of pile up effects which become important for the longer run. Estimating these effects are beyond the scope of this article and we only mention them as possible sources of uncertainties that will need to be accounted for in an experimental analysis.

\section{Summary}
\label{sec:concl}

The LHC has been running successfully and opening up new frontiers at the TeV scale. Among the many new possibilities for discovery at the LHC are color-octet resonances that are motivated in many BSM theories. These new colored particles will be produced copiously at a hadron collider and show up as resonance mass peaks in the dijet spectrum. Once these particles are discovered it is of paramount importance to measure the properties such as mass, spin and coupling structure to pinpoint the underlying theory. While some information (such as mass and spin) can be obtained from the discovery mode, other information vital to probing the theory such as chiral couplings cannot be obtained from that measurement. 

In this article we proposed a new channel, namely the associated production of a $W$ or $Z$ gauge boson with a color-octet resonance to provide information about the chiral structure. We combined the information from the dijet (discovery) mode with the associated production to uncover the chiral structure of the couplings of color-octet resonances to SM quarks. In order to make our study as widely applicable as possible, we have performed a phenomenological analysis of color-octets without being tied down to a specific theory. We sampled a wide range of masses, couplings and decay widths of color-octets and optimized the kinematic cuts to enhance the signal over the SM backgrounds. With this optimized analysis, we determined the sensitivity for the couplings and masses at the LHC and presented the results in the couplings plane for each mass after taking into account existing constraints. For one sample case of $m_C = 3\ \tev$ we also show the effect of being consistent with the $\afbt$ measurement at the Tevatron. 

We studied two scenarios at the LHC with c.m. energy of 14 TeV, one with the early data of 10 $\invfb$ integrated luminosity and the other with the longer run accumulating 100 $\invfb$ of data. As expected the early run is sensitive to lower masses and larger couplings but the reach dramatically improves both in terms of mass and couplings for the longer run. In particular the early run can probe masses up to 4.5 TeV but with relatively large couplings compared to the strong QCD coupling while the longer run explores a much larger region in couplings. We find encouraging results that the LHC will be able to provide information about the chiral structure for a wide range of couplings and masses and hence point us in the direction of the underlying theoretical structure.

\begin{acknowledgments}
We gratefully acknowledge Devin G. E. Walker for raising questions 
which led to this work, and for his collaboration on preliminary feasibility 
studies of a channel involving the top-quark.  This work is supported by the United States National Science Foundation under grants PHY-0854889 and PHY-0855561. We wish to acknowledge the support of the Michigan State University High Performance Computing Center and the Institute for Cyber Enabled Research. PI is supported by Development and Promotion of Science and Technology Talents Project (DPST), Thailand. 
\end{acknowledgments}

\bibliography{}

\end{document}